# Adapting cybersecurity frameworks to manage frontier AI risks

## A defense-in-depth approach

Institute for AI Policy and Strategy (IAPS)

October 13th, 2023

**AUTHORS**

Shaun Ee — Researcher

Joe O'Brien — Associate Researcher

Zoe Williams — Deputy Director

Amanda El-Dakhakhni - Acting Co-Director

Michael Aird - Acting Co-Director

Alex Lintz - Research Affiliate

# Executive Summary

**The complex and evolving threat landscape of frontier AI development requires a multi-layered approach to risk management ("defense-in-depth").** Developers and policymakers should ensure that "no single layer, no matter how robust, is exclusively relied upon" ([USNRC, 2021](#)). **Defense-in-depth is not a new concept, and is established in other high-stakes technological domains.** Among these, cybersecurity provides an especially rich set of materials for developers and policymakers to learn from. Drawing from cybersecurity, we identify three complementary approaches that frontier AI developers and policymakers can use to assess how comprehensive their risk management practices are and address significant gaps:

1. **Functional: Identify essential categories of safety and security activities ("functions")** that an organization must perform, and map these to a specified set of outcomes. This helps organizations to organize their risk management activities at a high level, and to assess if these activities are achieving the necessary outcomes. A functional approach is particularly helpful for identifying cross-cutting categories (e.g., organizational governance or insider security) that provide resilience against multiple known and unknown risks. It is also the most ready-to-adopt, based on the National Institute of Standards and Technology (NIST) AI Risk Management Framework (RMF) and supplementary guidance from other researchers that begins to adapt this framework to cover catastrophic risks from frontier AI. **We recommend that NIST or the Frontier Model Forum (FMF) establish consensus on the highest-priority categories of activities for frontier AI developers and develop a detailed catalog of measures ("controls") for these activities.**

2. **Lifecycle: Describe the frontier AI development lifecycle** and identify risk management activities that the organization must perform at each phase. This helps integrate safety and security into all stages of development, deployment, and monitoring. In cybersecurity, it has helped advance a "shift left" approach, i.e., designing safety into systems during development and tackling issues early in the software development lifecycle. While some AI development lifecycle frameworks exist, they need additional work to adapt to a frontier AI context and map appropriate risk management activities at each stage. **We recommend that the FMF develop a consensus model that captures these key activities for developers, and that AI developers, philanthropists, and government funders pursue research supporting a "shift left" for frontier AI safety and security.**

3. **Threat-based: Compile and describe the tactics, techniques, and procedures (TTPs) that threat actors use**, based on real-world evidence and what research suggests is possible. This approach typically focuses on TTPs to attack AI models ("effect on model"), but we suggest expanding it to cover TTPs *using* AI models ("effect on world"), given concerns about malicious use of AI. While MITRE ATLAS provides a solid starting point for "effect on model" TTPs, more work would be needed to build out both "effect on model" and "effect on world" approaches into comprehensive databases of TTPs and mitigations. **We recommend that MITRE, or the FMF, expand MITRE ATLAS into such a resource for frontier AI experts. We also recommend the US Cybersecurity and Infrastructure Agency (CISA) assess the effects of frontier AI systems on the top ten most vulnerable National Critical Functions.** Database owners should strongly consider limiting public access, due to the risk of facilitating attacks by malicious actors.

Below, we provide a summary capturing and comparing the main characteristics of these approaches.

| | Functional | Lifecycle | Threat-based |
|---|---|---|---|
| **Purpose** | Supports high-level risk management activities and resource allocation; assigns controls to lower-level outcomes. | Provides a holistic view of safety/security activities in software development, deployment, and operations. | Aids understanding of the motivations and methods of malicious actors in order to prepare effective defenses. |
| **Value proposition** | Ensures cross-cutting protections that provide resilience against known and unknown threats, even as technologies change. | Promotes "shift left" and "security by design"; calls attention to important deployment decisions and need for continuous monitoring. | Addresses adversarial actors; can facilitate whole-of-society defense against malicious use of frontier AI systems. |
| **Limitations** | Can be difficult to prioritize activities and evaluate coverage of threats. | Can omit cross-cutting categories of activities; does not focus on other actors' activities. | Exclusively focused on adversarial actors. |
| **Existing infrastructure that users can adopt for frontier AI risk management** | OK. Includes NIST AI RMF, CLTC risk profile for GPAI systems, and NIST CSF. | Limited. Various models exist, but no consensus and limited detail on specific activities. | Limited. MITRE ATLAS targets cyber experts, and has limited detail on frontier AI specific risks. |
| **Most suitable parties to conduct further research** | NIST, Frontier Model Forum | NIST, Frontier Model Forum | Frontier Model Forum, MITRE, CISA |

**To demonstrate how these frameworks can be useful, we conduct an illustrative gap analysis on the voluntary commitments made by leading AI labs in July 2023.** While these commitments are promising, applying the functional and lifecycle frameworks indicates that there is significant room for future commitments or legislation to fill, including: governance practices that facilitate a culture of risk management; pre-deployment review mechanisms tied to risk assessment results; and post-deployment monitoring and incident response.

Ultimately, we suggest that **frontier AI developers and policymakers first begin with a functional approach based on the NIST AI RMF and supplementary guidance,** as this approach is currently the most detailed of the three. However, these three approaches are complementary, as the complexity of AI systems and their threat landscape means that no single framework will satisfactorily capture all safety and security considerations. **We hence also recommend that key actors build out the lifecycle and threat-based approaches as described above.** Detailed recommendations are below.

## Recommendations

We identify actors who could implement these recommendations in the second column of the table below. These include the Cybersecurity and Infrastructure Agency (CISA), the Department of Defense (DoD), the Department of Energy (DOE), the Department of Health and Human Services (HHS), the Frontier Model Forum (FMF), MITRE, the National Institute of Standards and Technology (NIST), the National Science Foundation (NSF), the Partnership on AI (PAI), the UC Berkeley Center for Long-Term Cybersecurity (CLTC), and other general categories of actors such as frontier AI developers, researchers, and philanthropists.

| Functional | |
|---|---|
| **1 \| Establish consensus on which categories of activities in the NIST AI RMF are the highest priority for frontier AI developers.** (3.3.1 \| The NIST AI RMF) NIST and/or the FMF, with researcher input, should identify high-priority categories for frontier AI safety and security. To ensure defense-in-depth, frontier AI developers should implement multiple independent measures for these categories. | NIST (or FMF), with researcher input (e.g., CLTC) |
| **2 \| Develop a detailed catalog of measures ("controls") that are important for frontier AI safety and security.** (3.3.3 \| Providing detailed controls) For instance, NIST SP 800-53 lists 1,000 detailed controls for cybersecurity across 20 "families." No current equivalent exists for AI, and it would be useful for frontier AI developers to have a similar catalog focused on frontier AI safety and security. | NIST, or industry bodies like FMF or PAI |
| Lifecycle | |
| **3 \| Establish a detailed lifecycle framework for frontier AI that describes safety and security activities at each stage.** (4.3.2 \| Proposed lifecycle framework) This framework can build on work by the OECD while incorporating details from frontier AI developers, and should map activities to the NIST AI RMF where possible. It should ensure all phases are appropriately covered, which could include a "shift left" (see recommendation 4), and a stage for post-deployment monitoring and response. | FMF and/or NIST |
| **4 \| Pursue research that supports a "shift left" for frontier AI by emphasizing safety and security activities earlier in the development cycle.** (4.3.3.1 \| "Shifting left" on AI safety and security; 6.2.2 \| Lifecycle) Potential research areas could include: software requirement specification techniques borrowed from safety-critical domains, dataset curation techniques, and foundational research to build safer and more secure AI systems. | Frontier AI developers, philanthropists, and major government funders of AI R&D (e.g., the NSF, DoD, DOE, and HHS) |
| Threat-based | |
| **5 \| Restructure and expand MITRE ATLAS to further address attacks on frontier AI.** (5.3.2.1 \| An "effect on model" approach) MITRE ATLAS is a knowledge base of tactics, techniques, and procedures (TTPs) that malicious actors can use to attack AI systems. The high-level categories ("tactics") are | FMF, MITRE, and/or frontier AI developers |
| closely adapted from the equivalent cybersecurity knowledge base. We suggest restructuring these high-level tactics to reflect an AI-specific taxonomy (e.g., to include tactics like compromising training pipelines), and expanding on techniques and procedures that could enable misuse such as bypassing model guardrails. | |
| **6 \| Develop a common taxonomy of TTPs describing malicious use of frontier models to impact other actors and systems.** (5.3.2.2 \| <u>An "effect on world" approach</u>) The knowledge base should combine real-world evidence and what research suggests is possible. Database owners should strongly consider limiting public access, due to the risk of facilitating attacks by malicious actors. | FMF, MITRE, and/or frontier AI developers |
| **7 \| Establish a mechanism to assess and monitor potential effects of frontier AI systems on the top ten most vulnerable National Critical Functions.** (5.3.3 \| <u>Application to national critical functions</u>) These effects should be re-evaluated at least once every 1-2 years, and should be informed by the "effect on model" and "effect on world" databases described in recommendations 5 and 6. | CISA |

# Table of Contents

# 1 | Introduction

Frontier AI systems[1] have the potential to deliver significant benefits for society, but could also introduce catastrophic risks. For instance, members of Congress have proposed legislation to study how AI might facilitate the development of bioweapons and novel pathogens ([Heilweil, 2023](#)), a concern shared by industry executives who have testified to them ([Quach, 2023](#)). Other concerns from US officials revolve around the use of these models to plan cyberattacks or interfere with elections ([Vasquez, 2023](#)). OpenAI CEO Sam Altman and leading scientists have suggested that advanced AI could pose a risk on par with that of nuclear weapons, and that the existential threat it presents requires similar international regulation ([Roose, 2023](#); [Schmidt, 2023](#)).

These risks are complex, and many are still emerging. Most frontier AI systems are "general-purpose," which the EU AI Act defines as AI systems that "can be used in and adapted to a wide range of applications for which [they were] not intentionally and specifically designed" ([Hansen, 2023](#)). The general-purpose nature of these systems means that their capabilities, potential use cases, and associated risks are difficult to identify in full.

Because no single defense against such complex and evolving risks is foolproof, we recommend a "defense in depth" approach (*aka* "layered defense"). This is common in multiple high-stakes domains such as nuclear reactor safety, aviation, and cybersecurity.[2] Among these high-stakes domains, cybersecurity provides an especially rich and relevant set of materials for AI developers and policymakers to learn from.

We lay out three complementary risk management frameworks for frontier AI systems, adapted from cybersecurity: functional, lifecycle, and threat-based. We also illustrate how these can be used to evaluate current commitments such as the voluntary commitments by leading AI labs in July 2023 ([The White House, 2023a](#)), and provide concrete next steps for a variety of actors to adapt and implement these risk management frameworks to a frontier AI context. Together, this provides a pathway toward a risk management approach that can better ensure coverage of the wide variety of emerging threats faced in frontier AI.

---

[1] By "frontier AI systems," we use the Frontier Model Forum's definition as of October 2023: "large-scale machine-learning models that exceed the capabilities currently present in the most advanced existing models, and can perform a wide variety of tasks" ([OpenAI, 2023c](#)).
[2] See discussion at [Section 2](#) for further details.

# 2 | Defense-in-depth for frontier AI systems

Most definitions of defense-in-depth stem from the same principle: assembling multiple layers of defense so that "no single layer, no matter how robust, is exclusively relied upon."[3] But how can organizations apply this principle–i.e., how can they identify which layers to implement and whether they provide enough coverage of relevant risks? To answer this question, we borrow from defense-in-depth in other technological domains, including nuclear reactor safety and cybersecurity.

## 2.1 | Commonalities between domains implementing defense-in-depth

On January 23, 1961, the United States almost caused a nuclear detonation on its own soil. That day, a B-52 bomber was on a routine flight over Goldsboro, North Carolina when it went into an uncontrolled spin and began to break up. As it did so, it released two nuclear bombs, one of which began its detonation process ("US Plane in 1961 'Nuclear Bomb near-Miss,'" 2013).

The bomb had four safety mechanisms, of which one was disengaged due to the fall, and two of which were rendered ineffective by the aircraft breaking up. Only the final safety mechanism held fast. As a US nuclear engineer later wrote: "One simple, dynamo-technology, low voltage switch stood between the United States and a major catastrophe!" (_Goldsboro Revisited_, 2013; Rodriguez, 2019).

The Goldsboro accident demonstrates the hazards of relying on a single defense layer–or even just a handful of layers–for complex, high-stakes technological domains. In some such technological domains, like nuclear reactor safety and cybersecurity, practitioners often refer to defense-in-depth by name as an organizing principle for designing safe and secure systems.[4] Other such domains use similar layered defense concepts, though not always under the name of "defense-in-depth": e.g., the "Swiss cheese model" in aviation safety, "layers of prevention analysis" (LOPA) in chemical plant safety, the "web of prevention" in biosecurity, and the philosophy of "enhanced nuclear detonation safety" (ENDS) in nuclear weapon design.

The technological systems that use a defense-in-depth approach (or similar layered defense approaches) tend to share the following features: they have a high cost of failure, are complex to design and operate, and must manage an uncertain threat landscape.

---

[3] Adapted from the definition provided by the US's main regulatory body for nuclear power: "An approach to designing and operating nuclear facilities that prevents and mitigates accidents that release radiation or hazardous materials. The key is creating multiple independent and redundant layers of defense to compensate for potential human and mechanical failures so that no single layer, no matter how robust, is exclusively relied upon. Defense in depth includes the use of access controls, physical barriers, redundant and diverse key safety functions, and emergency response measures" (USNRC, 2021).

[4] For defense-in-depth in nuclear reactor safety, see (USNRC, 2021); for examples of defense-in-depth in cybersecurity, see May et al. (2006) or Information Technology Laboratory Computer Security Division (2021). As noted by Leveson (2020), defense-in-depth is an integral concept to nuclear reactor safety. By comparison, cybersecurity defense-in-depth coexists with multiple other concepts; see Appendix A-2 for discussion.

1. **High cost of failure:** System failures could lead to loss of life and/or large amounts of money, making it worthwhile to pay significant costs to minimize the chance of failure. Heavy investment to prevent potential loss of life is a common theme for safety-critical domains like aviation, biosecurity, chemical plant safety, nuclear power, and nuclear weapons. However, defense-in-depth is also common in other domains where economic damage predominates, such as in cybersecurity for digital systems that do not have physical actuators (e.g., financial systems).

2. **Complexity of technological system:** Systems have many parts which interact, are hard to track, and may require domain experts to understand and/or operate them. Failures could affect any of these parts, and result from a large number of causes or interactions between them.[5] Thus, an effective approach involves implementing defense layers that cover multiple parts and risks (breadth) in multiple ways (depth), rather than just strengthening the first line of defense.

3. **Uncertainty around adverse events:** Systems face a wide range of adverse events, which vary in terms of likelihood and magnitude, and whose impact on the system is difficult to predict. Adopting a defense-in-depth strategy helps compensate against uncertainty by providing buffer room and mitigating against the risk that adverse events cause multiple defenses to simultaneously fail.[6]

Most AI systems are already highly complex, and can fail or be attacked in unpredictable ways. Even before today's large language models (LLMs), researchers had already raised concern about how the "black-box" decision-making of AI systems made it difficult to anticipate, prevent, or learn from their failures,[7] and about how modern AI systems are vulnerable to attacks that are difficult for humans to discover.[8] Frontier AI models face additional challenges: for example, more powerful models may display unexpected dangerous capabilities (e.g., clearly explaining how to create a bioweapon when asked) that developers would not have predicted from previous model versions (Anderljung et al., 2023, pp. 10–13).

While current LLMs are not yet capable of causing widespread damage, these failures and attacks could become increasingly costly as capabilities progress and AI systems become more widely used. For example, Anthropic CEO Dario Amodei has testified to the US Senate that within three years, frontier models could be sufficiently capable enough for malicious actors to develop bioweapons using them (*Frontier Threats Red Teaming for AI Safety*, 2023; Quach, 2023).

---

[5] Reason (2000) argues that failure in these domains consists of one direct "active failure," behind which there are multiple background "latent failures." Major incidents occur when every defense layer is breached, which can expose all points of failure.

[6] From Drouin et al. (2016) at p. 202: "There is general agreement that defense-in-depth is needed to compensate for uncertainties," which include uncertainties about design and operation, the performance of the plant under various conditions, how adverse events will progress, etc.

[7] In a famous paper from 2016, researchers showed graduate students an AI model that apparently could differentiate between wolves and huskies. Less than half were able to figure out that the model actually had not "learned" what a wolf was–instead, it was ignoring the animal to look at the background, classifying the image as a wolf solely if snow was present (M. T. Ribeiro et al., 2016).

[8] For example, adding a subtle pattern of noise to an image of a panda can fool an image classification algorithm into misclassifying it as a gibbon, even if there is no apparent difference to a human (Heaven, 2019).

## 2.2 | Defense-in-depth in nuclear power

Nuclear power is the industry that uses defense-in-depth most prominently, and has codified several principles that make it a useful starting point for understanding defense-in-depth.[9] However, there are also several key differences between nuclear reactor systems and AI systems that limit the direct applicability of frameworks in nuclear reactor safety to frontier AI systems.

Defense-in-depth in nuclear power is built around the assembly of sequential layers to limit the progression of an incident. One group convened by the International Atomic Energy Agency (IAEA) lays out five "levels of defense" that escalate from preventing deviations to preventing core damage to mitigating the fallout of an accident.[10] (See Appendix A-1 for a full version of these levels.) This ensures that even if an incident begins to occur, there are still safeguards in place to slow or halt the incident's progression.

A nuclear power defense-in-depth strategy also commonly involves several additional principles:[11]

- **Independence:** Layers must be *independent* of each other, so that "common cause failures" do not overcome multiple layers at once. An example of such a common cause failure was the tsunami that caused the Fukushima Daiichi nuclear power plant disaster, which disabled several sources of power simultaneously: external power lines, emergency generators, and several backup batteries. To ensure that these defense layers did not fail simultaneously, the plant owner should have waterproofed some of them or moved them to higher ground (Hibbs & Acton, 2012).

- **Diversity:** Layers must be *diverse* in their design, to increase coverage and hedge against risks that come with using specific safety measures.[12] For example, plant operators can use different equipment designs or manufacturers for safety mechanisms, in case one is susceptible to an unidentified problem.

- **Redundancy:** Layers should be *redundant*, using backups to minimize the risk of multiple failures.

These general principles are useful to consider for frontier AI systems, but many specific measures used in a nuclear power defense-in-depth context do not map over well to frontier AI systems. As other researchers have written, the defense-in-depth approach used in nuclear power is a specific engineering approach that gained popularity in the industry due to unique historical reasons and the physical properties of nuclear power plants, particularly the ability to revert to a "safe state" (i.e., shutting down the plant) (Leveson, 2020, pp. 19–20). Dissimilarities include:

---

[9] For a detailed history of defense-in-depth in the nuclear power industry, see (Drouin et al., 2016).

[10] From International Nuclear Safety Advisory Group (1996) Table 1 (p. 6) on "Levels of Defense in Depth." (The author of this publication, the International Nuclear Safety Advisory Group or INSAG, is convened by the IAEA.) Although note that per Drouin et al. (2016), p. 207, "there is no agreement in the number of layers of defense. They vary from two layers, prevention and mitigation, to five layers" depending on the national or international agency consulted.

[11] For example, these three principles are laid out in Drouin (2016).

[12] Diversity and independence are closely related, but "diversity" in nuclear plant safety often refers more narrowly to design features–e.g., motor-driven vs. steam-powered pumps (Drouin, 2016).

- **Frontier AI systems are software systems, while nuclear reactors are hardware systems.** Some principles like hardware redundancy are hence less applicable to frontier AI systems, and it is harder to assess the probability of failure of frontier AI safety mechanisms.[13]

- **Frontier AI systems have many possible design architectures and failure modes, while nuclear reactors and incidents follow a more limited number of archetypes.** Nuclear plants typically fall into a relatively small number of well-characterized designs.[14] Accidents also typically have a standard format, progressing first to a core meltdown, and then to dispersal of radiological material. In domains like cybersecurity and AI, systems and pathways to failure are much more varied.

Defense-in-depth in nuclear reactor safety is still a powerful motivating example for frontier AI safety and security, and provides some useful principles to follow. However, these dissimilarities mean that it is not a good model to copy directly for frontier AI. Instead, we suggest cybersecurity–another domain that frequently uses defense-in-depth, and which illustrates the tension between high resource demands and the need for agility in the face of uncertainty.

## 2.3 | Cybersecurity as a model for AI

### 2.3.1 | Cybersecurity defense-in-depth in the 2000s and beyond

Cybersecurity practitioners including the US Department of Defense (DoD) and Carnegie Mellon University Software Engineering Institute (CMU SEI) have advocated for using a defense-in-depth approach since the 2000s,[15] but the actual interpretation of defense-in-depth varies considerably in cybersecurity. Unlike in nuclear reactor safety, where defense-in-depth strategies have coalesced around mitigating the damage at each stage of an accident, cybersecurity practitioners use "defense-in-depth" in several overlapping ways.

Defense-in-depth approaches in cybersecurity largely differ in terms of how they categorize types of security controls.[16] Some popular approaches adopt very general high-level categories, such as "physical,

---

[13] For example, hardware behaves in a predictable (i.e., deterministic) way, making it possible for engineers to assess hardware safety using methods that estimate the rate of failure. This is not true of software safety, because software often behaves non-deterministically, and software failures are systematic (i.e., they arise from issues to do with the system as a whole, rather than failures of individual components) (Khlaaf, 2023, pp. 10–11).

[14] "Nuclear power engineering has, until recently, used a few designs for which a lot of past experience can be accumulated and has been very conservative about introducing new technology, such as digital systems. That conservatism is giving way to greatly increased use of digital instrumentation and control" (Leveson, 2020, p. 19).

[15] E.g., see May et al. (2006) for a 2006 paper on defense-in-depth by CMU SEI. Also see DoD Directive 8500.01, issued in October 2002, which defines defense-in-depth as "The DoD approach for establishing an adequate IA posture in a shared-risk environment that allows for shared mitigation through: the integration of people, technology, and operations; the layering of IA solutions within and among IT assets; and, the selection of IA solutions based on their relative level of robustness" (US Department of Defense, 2002, p.18).

[16] I.e., safeguards and countermeasures to protect information (*Security Control*, n.d.).

technical, administrative," or "people, process, technology."[17] Others reflect the need for attackers to penetrate successive network and system layers to gain access to an organization's "crown jewels": for example, perimeter, network, endpoint, application, data ([Jacobi, 2023](#)). Yet others focus on identifying families of controls that fulfill certain functions, such as an approach by CMU SEI that splits controls into eight categories: compliance, risk, identity, authorization, accountability, availability, configuration, and incident management ([May et al., 2006](#)). (For a brief review of these approaches, see [Appendix A-2](#).[18])

Despite these varied approaches, defense-in-depth has persisted as a commonly referred-to principle in cybersecurity for at least twenty years.[19] We believe the key lesson from this period is that multiple approaches to structuring a defense-in-depth approach are necessary for fields like cybersecurity and AI, where systems are complex and the threat landscape is uncertain. We build on this in the rest of the paper by suggesting multiple complementary approaches that frontier AI developers and policymakers can use to achieve an appropriate coverage of risks.

## 2.3.2 | Complementary approaches to address evolving capabilities and threats

The significant variation in cybersecurity frameworks reflects that in cybersecurity, as in frontier AI, there is a need for multiple perspectives to address the large range of evolving threats. While it is easy to see the lack of a single robust cybersecurity defense-in-depth framework as an industry-wide failure, we argue that it in fact reflects the dynamic and uncertain nature of cybersecurity as a domain. Three features–changing systems, changing threats, and determined adversaries–make this variety of approaches necessary in both cybersecurity and frontier AI.

**First, diverse and rapidly evolving architectures and capabilities** are a hallmark of both cybersecurity and frontier AI. Cybersecurity defenders must cover everything from the financial sector to the energy grid, and technological change over time has forced large strategic shifts. For example, the shift from perimeter defense to zero-trust was catalyzed by the rise of cloud computing, the increasing use of non-company-issued personal devices, and the emergence of remote work ([Rose et al., 2020](#)). Frontier AI systems have also evolved rapidly, and their pace of change may accelerate.[20] Current AI policy discussions are dominated by large language models (LLMs) like OpenAI's GPT-4, and other multimodal AI systems that combine computer vision, language, and physical actuators (*[RT-2, n.d.](#)*). But seven years ago (in 2016), the state-of-the-art was DeepMind's game-playing AlphaGo, built on a technique known as reinforcement learning (*[From AI to Protein Folding, n.d.](#)*); the underlying concept for current LLMs, the "transformer," was only developed in 2017 ([Vaswani et al., 2017](#)).

---

[17] NIST SP 800-53, for example, defines defense-in-depth as "An information security strategy that integrates people, technology, and operations capabilities to establish variable barriers across multiple layers and missions of the organization" ([Joint Task Force, 2020](#)). Other contemporary sources like [Fruhlinger (2022)](#) and [Stewart et al. (2015)](#) use the "physical, technical, administrative" taxonomy.

[18] Some practitioners frame other cybersecurity concepts, like zero-trust and "assume breach," as opposed to specific versions of defense-in-depth. However, we use defense-in-depth more broadly in the sense of "layered defense," and so view these concepts as complementary. For additional discussion, see [Appendix A-2](#).

[19] For example, recent publications like [Stewart et al. (2015)](#) and [Information Technology Laboratory Computer Security Division (2021)](#) continue to refer to defense-in-depth.

[20] Researchers at [Epoch](#) have investigated historical trends in AI progress, in the expectation that these could inform future rates of AI progress. At the time of publication, these trends included training compute growth at a rate of 4.2x per year, and algorithmic improvements resulting in 2.5x less physical compute needed for the same image classification each year ([Epoch, 2023](#)).

This pace of change is faster than domains like nuclear reactor safety or aviation, where hardware and software designs are slow to change, and engineers are conservative about adopting new technologies. Such changes also make defense-in-depth strategies based on particular system or network architectures less likely to hold up over time; for example, defense-in-depth interpretations rooted in strong perimeter defense and protecting successive network zones have required supplementing with zero-trust approaches due to the changing environment, as previously mentioned.

**Second, there is uncertainty about the scope and type of vulnerabilities and incidents.** In cybersecurity, about 20,000 new vulnerabilities in common software products are disclosed annually, requiring defenders to triage these by assessing which are critical and must be addressed immediately.[21] Frontier AI researchers are also continually identifying new vulnerabilities and classes of vulnerabilities, such as methods to automatically generate sequences that bypass LLM safety guardrails (i.e., measures that prevent the output of harmful or toxic content) (Claburn, 2023).[22] Also, both cybersecurity and AI researchers must guard against a diverse range of possible incidents; malicious actors in cybersecurity can conduct attacks designed to steal data, corrupt or destroy it, or deny legitimate users access to it,[23] while senior government officials have raised concern about frontier AI systems causing systemic financial sector issues (Gensler & Bailey, 2020; Sorkin et al., 2023), or being used to conduct disinformation campaigns, facilitate cybercrime, and produce biological and chemical weapons.[24]

**Third, defenders face a wide range of adversaries that are actively trying to exploit their weak points.** These include attacks from well-resourced and sophisticated attackers, as well as a large number of less-resourced attacks–a sharp difference from many other defense-in-depth domains where the threat is more static. In cybersecurity, Russia-linked actors executed one of the largest breaches of the last decade–the SolarWinds attack–by using a cybersecurity software provider's update process to bypass standard defenses and infect 18,000 companies (Herr et al., 2021; *SolarWinds Compromise, Campaign C0024,* *2023*). This attack was sufficiently skilled that three years later, experts are still not confident that the attack's full extent has been uncovered (Zetter, 2023).

Similarly, adversarial actors will likely find creative ways to attack frontier AI, and to use it maliciously. Previous incidents show that highly accessible AI systems are likely to attract interest from members of the public interested in breaking their safeguards; for example, in 2016, Twitter users were able to induce Tay, a Microsoft chatbot, to produce highly offensive content within 24 hours of it going live (Vincent, 2016). More considered and deliberate attacks could cause catastrophic harm, such as using advanced AI systems to produce biological or chemical weapons (*Frontier Threats Red Teaming for AI Safety*, 2023;

---

[21] The Redscan Team (2021): "There have been more security vulnerabilities disclosed in 2021 (18,439)* than in any other year-to-date – averaging more than 50 CVEs logged each day." Tenable (2018): "Our research shows that enterprises must triage more than 100 critical vulnerabilities a day... In 2017 alone, an average of 41 new vulnerabilities were published daily – that's 15,038 for the year."

[22] Prominent technologist Bruce Schneier has said of security research into current AI models, "This is computer security 30 years ago. We're just breaking stuff left and right" (Bajak, 2023).

[23] One way of conceptualizing cybersecurity incidents is the "CIA triad" (confidentiality, integrity, and availability). For example, a destructive attack like website defacement might affect the integrity (i.e., correctness) and availability of an organization's data, while the misconfiguration of a cloud asset to be Internet-accessible when it should not be could affect the confidentiality of data instead (Office of Information Security, n.d.).

[24] See The White House (2023a) and Quach (2023) for discussion of biological and chemical weapons; see Vasquez (2023) for discussion of disinformation and cyberattacks.

[Urbina et al., 2022](#)). In both cybersecurity and frontier AI, the sheer quantity of possible attackers means that defenders must put considerable effort into covering all their bases.

Because of this range of uncertain and evolving risks, cybersecurity has developed a variety of frameworks that look at defense from different angles. For instance, a threat-based approach is valuable for breaking down the best actions to take for defending against threat actors who are using known techniques, but does not hold up as well in providing protection against unknown risks. A functional approach partially compensates for this by suggesting functions such as strong governance, access controls, and monitoring, which can help organizations manage risks even if they have never been seen before. Frontier AI also faces all the challenges explored above, and therefore should also implement multiple frameworks in order to achieve a strong defense against known, anticipated, and unknown risks.

### 2.3.3 | Benchmarking measures to the appropriate level of risk

Users should adopt the correct frame of reference when adapting cybersecurity frameworks–for example, a high-risk AI system with national security implications will require more extensive measures than lower-risk systems used by small and medium enterprises. While current frontier AI systems such as GPT-4 are unlikely to have severe economic or national security impacts ([OpenAI, 2023a](#)), future AI systems plausibly could. Anthropic CEO Dario Amodei has testified before the US Senate, for example, that frontier models could plausibly be capable enough that malicious actors could use them to develop bioweapons within the next three years (*[Frontier Threats Red Teaming for AI Safety, 2023](#)*; [Quach, 2023](#)). Additionally, some researchers have suggested that sophisticated systems could pose an existential threat to humanity if they pursue their own subgoals ([Brown, 2023](#)).

The robustness and costliness required of defense-in-depth measures for some frontier AI systems might hence be closer to the serious cybersecurity measures used for national security or critical infrastructure, such as air-gapping, the use of Sensitive Compartmented Information Facilities (SCIFs), or other measures laid out in NIST Special Publication 800-172 (NIST SP 800-172). (See [Appendix A-2](#) for details on NIST SP 800-172.) Frontier AI developers and policymakers may also wish to look to other domains besides cybersecurity for inspiration. Nuclear weapons, for example, rely on an extensive series of failsafes and security measures to prevent unauthorized or accidental use.

A tiered approach to risk assessment will be needed to balance safety and innovation for frontier AI systems. For example, frontier AI company Anthropic has laid out a "Responsible Scaling Policy" (RSP) that describes the safety measures required for AI systems at different levels of capability (*[Anthropic's Responsible Scaling Policy, 2023](#)*).[25] Inspired by the biosafety level (BSL) standards used to manage the handling of dangerous pathogens, this policy lists at least four tiers of systems at different "AI safety levels" (ASLs):

- ASL-1: Smaller, narrower models (e.g., large language models from 2018, or chess-playing systems)
- ASL-2: Current large models (e.g., large language models from 2023)
- ASL-3: Future models with significantly higher risks (e.g., AI systems that could make malicious acts such as bioterrorism or cyberattacks substantially easier, or that display limited autonomy)

---

[25] Another organization, the Alignment Research Center, has also suggested more general desiderata for RSPs ([ARC Evals, 2023](#)).

- ASL-4 and beyond: Speculative future models with extremely high risks

As AI capabilities progress, frontier AI developers and policymakers should establish consensus around the levels of the ASL system, or an equivalent tiered system, and the measures that frontier AI developers will have to adhere to at each level.

## 2.4 | Three approaches to AI defense-in-depth

How can frontier AI developers and policymakers implement a defense-in-depth strategy to address potential catastrophic risks? In the remainder of this paper, we propose three complementary approaches that can inform such a strategy for frontier AI: a functional, lifecycle, and threat-based approach. We have ordered these approaches from the easiest to adopt (i.e., less technical, and more existing resources) to the hardest to adopt (i.e., most technical, and less existing resources).

- **Functional: Identify essential categories of safety and security activities ("functions")** that an organization must perform, and map these to a specified set of outcomes. This helps organizations to organize their risk management activities at a high level, and to assess if these activities are achieving the necessary outcomes.

  We view this approach as valuable because it identifies outcomes (e.g., a deployed AI system having been demonstrated to be valid and reliable) and cross-cutting measures (e.g., organizational governance or insider security) that will remain relevant even as AI capabilities and threats evolve. Because of this outcome-driven aspect, a functional approach should (if applied well) help users avoid the trap of "checklist compliance."

  This approach has the best-developed infrastructure of the three we describe. The NIST AI Risk Management Framework is a detailed resource for a functional approach that is widely viewed as authoritative in its field.

- **Lifecycle: Describe the AI development lifecycle, and identify important safety and security activities that the organization must perform at each phase.** This approach can provide a holistic view of such activities across software development, deployment, and operations.

  We view this approach as valuable because it helps promote a "shift left" and "security by design" approach–i.e., addressing safety and security issues early in the development lifecycle, rather than waiting until the end to test and mitigate. It also calls attention to critical decision-making nodes in the deployment process, and underscores the importance of monitoring and response in post-deployment system operation.

- **Threat-based: Compile and describe the tactics, techniques, and procedures (TTPs) that threat actors use**, based on real-world evidence and what research suggests is possible. Existing resources focus on TTPs used to attack AI models ("effect on model"), but we suggest expanding it to cover TTPs *using* AI models ("effect on world"), given concerns about malicious use of AI.

  We view this approach as valuable because frontier AI facilitates potentially catastrophic risks (e.g.,

the creation of bioweapons) from a wide range of threat actors including non-state actors, nation-states, and sophisticated goal-directed AI systems. In addition to helping frontier AI developers identify and evaluate defenses to these risks, a comprehensive database of TTPs involving frontier AI systems could also facilitate a whole-of-society approach that involves developing countermeasures to malicious use of such systems, and strengthening societal resilience.

Our aim is not to litigate the interpretation of defense-in-depth in cybersecurity, but rather to ask the question: if one had to build a defense-in-depth strategy for frontier AI, what are the most useful cybersecurity constructs one could borrow?[26] We chose these approaches to provide complementary perspectives that frontier AI developers and policymakers can use to identify gaps and weak points in their defenses. We see this as a necessary prerequisite to deciding which defenses (or types of defenses) perform the most critical role(s), and therefore where having layered independent defenses is most necessary. In some ways, one can frame this as having defense-in-*breadth* as a prerequisite to defense-in-depth.[27]

We also point towards existing resources that suggest which of these defenses might be most critical (in cases where this analysis exists), and highlight future research priorities to extend on this selection of high-priority categories and the guidance for their implementation. We discuss further guidelines for use in Section 6.

# 3 | Functional approach

NIST's Cybersecurity Framework (CSF) and AI Risk Management Framework (AI RMF) exemplify a commonly used approach to risk management that involves grouping relevant activities into risk management "functions." For example, the CSF has five functions: "Identify, Protect, Detect, Respond, Recover," while the AI RMF has four functions: "Govern, Map, Measure, Manage." These functions can facilitate high-level risk management strategy and decision-making.[28] For frontier AI developers, their main advantage is that they can be used to identify cross-cutting measures (e.g., organizational governance or insider security) that provide resilience against a variety of risks even when capabilities and threats change.

The detailed and consensus-driven AI RMF makes the functional approach the most mature of our three recommended approaches, but additional work is needed to tailor the AI RMF to frontier AI systems. The AI RMF is meant for "organizations of all sizes and in all sectors and throughout society" ([Tabassi, 2023, p. 2](#)), and includes other aspects of AI trustworthiness besides safety and security (e.g., privacy). Both the AI RMF and CSF provide a baseline for sector-specific risk profiles and frameworks to build on—for example, in cybersecurity, NIST and industry actors have built risk profiles on top of the CSF to cover manufacturing and election security. For frontier AI, more targeted guidance will similarly be needed. For example, supplementary guidance by researchers from the UC Berkeley Center for Long-Term Cybersecurity (CLTC) identifies high-priority activities for frontier AI developers to reduce catastrophic AI risks.

---

[26] We have steered away from some common cybersecurity frameworks explicitly labeled as "defense-in-depth" (e.g., three-factor approaches like "physical, technical, administrative") because they are not detailed enough to inform a robust strategy. Further discussion is available at Appendix A-2.
[27] NIST SP 800-53 Rev. 5, for example, explicitly identifies the second approach we suggest (the lifecycle approach) as a defense-in-breadth approach: that is, they define defense-in-breadth as "A planned, systematic set of multidisciplinary activities that seek to identify, manage, and reduce risk of exploitable vulnerabilities at every stage of the system, network, or subcomponent life cycle, including system, network, or product design and development; manufacturing; packaging; assembly; system integration; distribution; operations; maintenance; and retirement." (NIST CSRC, n.d.) However, we avoid using this definition because it is not common in industry.
[28] NIST describes the CSF as "aiding organizations in easily expressing their management of... risk at a high level and enabling risk management decisions" ("The Five Functions," 2018).

Frontier AI developers can use the AI RMF and such supplementary guidance to achieve a defense-in-depth approach by identifying the highest-priority categories of activities within the AI RMF for safety and security, and implementing multiple independent measures ("controls") for these high-priority categories. In addition to targeted risk profiles like the CLTC guidance, this will also require more research to establish a catalog of safety and security controls for frontier AI. For example, in cybersecurity, NIST SP 800-53 lists over 1,000 cybersecurity controls ([Joint Task Force, 2020](#)); no similar catalog exists for frontier AI, which makes it more difficult for frontier AI developers and policymakers to identify potential measures, evaluate the completeness of existing measures, and compare frontier AI developers' approaches.

## 3.1 | What does this look like in cybersecurity?

The NIST Cybersecurity Framework (CSF) is a commonly used risk management framework that illustrates the functional approach. As of October 2023, NIST CSF version 1.1 covers five functions: Identify, Protect, Detect, Respond, and Recover (IPDRR).[29] These top-level categories of activities "aid organizations in easily expressing their management of... risk at a high level and enabling risk management decisions" (["The Five Functions," 2018a](#)). The CSF also breaks these top-level categories down into smaller subcategories and links them to relevant guidance, allowing practitioners to analyze their organization's coverage of risk management activities in detail.

We do not explain the NIST CSF in detail because an AI-focused equivalent of the CSF already exists (i.e., the NIST AI RMF; see Section 3.3 below). Instead, we suggest frontier AI researchers focus directly on how to expand on the AI RMF, but [Appendix A-3](#) contains further details on the CSF if useful.

One useful lesson from the CSF is that targeted standards and guidance are needed to support the CSF and AI RMF in addressing catastrophic risks. Like the AI RMF, the CSF does not focus on catastrophic risks—e.g., risks that could cause a large loss of human life or economic value, or have a significant impact on society. The CSF instead functions as a general risk management framework for organizations of all missions and sizes, though it was initially created to address cybersecurity risks to US critical infrastructure.[30] Frontier AI developers adapting functional cybersecurity frameworks for AI should also consider other standards that are more specialized and have stricter requirements than the NIST CSF.

---

[29] NIST CSF Version 2.0, to be re-released in early 2024, will add a new "Govern" function (*NIST Cybersecurity Framework 2.0 Concept Paper: Potential Significant Updates to the Cybersecurity Framework, 2023, p. 10*).

[30] "While the CSF was originally developed to address the cybersecurity risks of critical infrastructure first and foremost, it has since been used much more widely" (*NIST Cybersecurity Framework 2.0 Concept Paper: Potential Significant Updates to the Cybersecurity Framework, 2023, p. 4*).

- "Framework profiles" based on the NIST CSF adapt it for specific industries and use cases, such as election infrastructure or manufacturing (NIST, 2021). Because the NIST CSF is very broad and covers many possible categories of activities for many actors, risk profiles help prioritize the categories of activities most important to mission objectives common to the industry or use case (e.g., "Process and Maintain Voter Registration," or "Maintain Human Safety").[31] These profiles can also then prescribe additional guidance, and possibly tailor this to subcategories (e.g., "high-impact" manufacturing systems) (Stouffer et al., 2020, pp. 17-45).

- Other cybersecurity standards, such as NIST Special Publication 800-172 (NIST SP 800-172), focus on more stringent requirements for organizations to protect themselves against sophisticated nation-state cyberattacks. NIST SP 800-172 has three main thrusts, intended to counter sophisticated adversaries: (1) penetration-resistant architecture, (2) damage-limiting operations, and (3) designing for cyber resiliency and survivability.[32] (Details at Appendix A-2.)

## 3.2 | Why take a functional approach?

The functional approach is useful for frontier AI because:

- **Implementing cross-cutting measures can provide resilience against both known and unknown risks.** The CSF and AI RMF include categories of activities that do not fit easily into specific software development lifecycle phases and threat models, such as identifying roles and responsibilities, establishing oversight processes, and improving awareness and training. Strengthening these measures can be helpful even as technological capabilities, risks, and threat actors change.

- **The AI RMF is relatively mature compared to other AI risk management resources, and covers a comprehensive range of activities in substantial detail.** This allows users to identify high-priority categories of activities, and to evaluate their own risk management programs against the AI RMF at a granular level.

The NIST CSF is also used for other purposes, such as providing a high-level summary of risk management efforts, or overviewing spending on risk management. See Appendix A-3 for details.

## 3.3 | Usage for frontier AI governance

To illustrate a function-based approach to governing frontier AI systems, we refer to the "Govern, Map, Measure, Manage" framework described in the NIST AI Risk Management Framework (NIST AI RMF), which provides a comprehensive description of the categories of defenses. While the NIST AI RMF is a valuable resource, it requires tailoring for frontier AI development, for which we reference guidance from UC Berkeley's Center for Long-Term Cybersecurity (CLTC) that identifies selected categories as high-priority. Lastly, we suggest that a defense-in-depth approach can be achieved by providing multiple independent layers for each high-priority category; while some such guidance for doing so already exists, we suggest that NIST or other researchers could further build out a catalog of controls for frontier AI safety and security.

---

[31] For example, p. 8-12 from the risk profile on manufacturing: Stouffer et al. (2020); or pp. 14-46 from the risk profile on election infrastructure: Brady et al. (2021).

[32] "The enhanced security requirements provide the foundation for a multidimensional, defense-in-depth protection strategy through (1) penetration-resistant architecture, (2) damage-limiting operations, and (3) designing for cyber resiliency and survivability that support and reinforce one another" (Information Technology Laboratory Computer Security Division, 2021).

### 3.3.1 | The NIST AI RMF

The NIST AI Risk Management Framework (RMF) is intended as a voluntary framework for "organizations of all sizes and in all sectors and throughout society," and is agnostic as to use case and sector (Tabassi, 2023, p. 2). After Congress directed NIST to draft the AI RMF in the National Artificial Intelligence Initiative Act of 2020, NIST undertook an 18-month drafting period that included extensive feedback from government, industry, and civil society, and released the first complete version of the AI RMF in January 2023 (Rep. Johnson, 2020).[33] We focus on the NIST AI RMF because this extensive input process, and NIST's prominence as a major standards-setting organization, make the AI RMF a detailed, credible, and consensus-driven resource for organizations looking for guidance on how to manage AI risks.

The NIST AI RMF is organized around four functions, representing high-level categories of activities:

- **Govern:** A culture of risk management is cultivated and present.
- **Map:** Context is recognized and risks related to context are identified.
- **Measure:** Identified risks are assessed, analyzed, or tracked.
- **Manage:** Risks are prioritized and acted upon based on a projected impact (Tabassi, 2023, p. 20)

Like the CSF, the AI RMF then breaks these functions down into more detailed subcategories and guidance (see Appendix B for more details).

### 3.3.2 | Tailoring the AI RMF to frontier AI safety and security concerns

The NIST AI RMF does not focus on the safety and security of frontier AI models. It is intended as a resource for a broad audience, covering a wider range of actors (e.g., developers of smaller AI models, downstream users)[34] and concerns (e.g., privacy and other aspects of trustworthiness) (Tabassi, 2023, pp. 12-18). Given that the US government, frontier AI developers, and leading scientists have raised concerns about catastrophic risks from frontier AI models (Brown, 2023; The White House, 2023b), a specialized framework is needed to complement the NIST AI RMF.

Several efforts are underway to establish supplementary guidance related to the NIST AI RMF for frontier AI systems and large language models. NIST is currently developing a risk profile focusing on generative AI, supported by a public working group announced by the White House (the NIST GAI

---

[33] See NIST Information Technology Laboratory (2023) for timeline of NIST AI RMF drafting.
[34] "The goal of the AI RMF is to offer a resource to the organizations designing, developing, deploying, or using AI systems to help manage the many risks of AI and promote trustworthy and responsible development and use of AI systems" (Tabassi, 2023, p. 2).

PWG).[35] The GAI PWG is expected to address four aspects of generative AI: governance, content provenance, pre-deployment testing, and incident disclosure ([NIST AIRC Team, n.d.-a](#)). Another group of researchers at UC Berkeley's Center for Long-Term Cybersecurity (CLTC) is developing supplementary guidance to the AI RMF that is focused on catastrophic AI risks ([Barrett, Hendrycks, et al., 2023](#)).

Because the CTLC guidance is currently the most detailed set of recommendations describing best practices for addressing potential catastrophic risks from frontier AI systems, we focus on the CLTC guidance for our discussion. As of October 2023, the most recent draft of the CLTC framework recommends that frontier AI developers treat the following points as highest-priority ([Barrett, Newman, et al., 2023](#)):

---

**High-priority categories of activities identified by CLTC supplementary guidance**

- **Take responsibility for risk assessment and risk management tasks for which your organization has substantially greater information and capability than others in the value chain** (Section 3.1, Govern 2.1)
- **Set risk-tolerance thresholds to prevent unacceptable risks** (Map 1.5)
- **Identify the potential uses, and misuses or abuses for a general purpose AI system (GPAIS),** and **identify reasonably foreseeable potential impacts** (e.g., to fundamental rights) (Map 1.1)
- **Identify whether a GPAIS could lead to significant, severe or catastrophic impacts,** e.g., because of correlated failures or errors across high-stakes deployment domains, dangerous emergent behaviors, or harmful misuses and abuses by AI actors (Map 5.1)
- **Use red teams and adversarial testing** as part of extensive interaction with GPAIS **to identify dangerous capabilities, vulnerabilities** or other emergent properties of such systems (Measure 1.1)
- **Track important identified risks** (e.g., vulnerabilities from data poisoning and other attacks or objectives mis-specification) even if they cannot yet be measured (Measure 1.1 and Measure 3.2)
- **Implement risk-reduction controls as appropriate** throughout a GPAIS lifecycle, e.g., independent auditing, incremental scale-up, red-teaming, and other steps (Manage 1.3, Manage 2.3, and Manage 2.4)
- **Incorporate identified AI system risk factors, and circumstances that could result in impacts or harms, into reporting to internal and external stakeholders** (e.g., to downstream developers, regulators, users, impacted communities, etc.) on the AI system as appropriate, e.g., using model cards, or system cards (Govern 4.2)
- **Check or update, and incorporate, each of the above when making go/no-go decisions,** especially on whether to proceed on major stages or investments for development or deployment of cutting-edge large-scale GPAIS (Manage 1.1)

---

[35] [Biden-Harris Administration Announces New NIST Public Working Group on AI" (2023)](#): "The Public Working Group on Generative AI will... help NIST develop key guidance to help organizations address the special risks associated with generative AI technologies... it will serve as a vehicle for gathering input on guidance that describes how the NIST AI Risk Management Framework (AI RMF) may be used to support development of generative AI technologies. This type of guidance, called a profile, will support and encourage use of the AI RMF in addressing related risks."

### 3.3.3 | Providing detailed controls

To achieve defense-in-depth, frontier AI developers should implement multiple independent measures for each of these high-priority categories, which reduces the risk of any single failure leading to a catastrophic outcome. To identify such measures, frontier AI developers can draw on existing guidance–for example, NIST provides an online "playbook" alongside the AI RMF that includes suggested actions for users (NIST AIRC Team, n.d.-b), and the supplementary guidance by CLTC also includes specific actions and reference materials. They can also draw on issue-specific studies by researchers, such as an overview of risk assessment techniques by (Koessler & Schuett, 2023).

Ideally, having a comprehensive catalog of controls for frontier AI would make it easier for actors to identify potential measures, evaluate the completeness of existing measures, and compare frontier AI developers' approaches. In cybersecurity, NIST SP 800-53 is probably the most comprehensive such catalog, listing over 1,000 cybersecurity controls that are divided into 20 control "families," such as awareness and training, incident response, and supply chain risk management.[36] For example, control AC-6 ("least privilege," under the "access control" family) is: "Employ the principle of least privilege, allowing only authorized accesses for users (or processes acting on behalf of users) that are necessary to accomplish assigned organizational tasks." However, at this time there is no similarly comprehensive catalog of controls for frontier AI safety and security.

We recommend that an industry body or NIST draw up a catalog of controls addressing safety and security for frontier AI models, performing a similar role to NIST SP 800-53. This would ideally be driven by NIST, but an industry body focused on AI safety and security could potentially perform a similar role, such as the Frontier Model Forum, the Partnership on AI, or another information-sharing forum or mechanism as indicated in the voluntary commitments announced by leading AI companies and the White House in July 2023.[37]

In the interim, organizations should use existing standards and best practices for individual aspects of AI safety and security and curate these resources appropriately, e.g., using NIST SP 800-53 specifically for system cybersecurity, or existing resources applying the NIST AI RMF toward catastrophic risk management (Barrett, Newman, et al., 2023).

### 3.3.4 | Defense-in-depth using the NIST AI RMF

To summarize, using the NIST AI RMF and supplementary guidance such as the CLTC guidance, frontier AI developers can take the following steps to achieve an effective defense-in-depth approach:

---

[36] The twenty categories are: access control; awareness and training; audit and accountability; assessment, authorization, and monitoring; configuration management; contingency planning; identification and authentication; incident response; maintenance; media protection; physical and environmental protection; planning; program management; personnel security; personally identifiable information processing and transparency; risk assessment; system and services acquisition; system and communications protection; system and information integrity; and supply chain risk management (Joint Task Force, 2020).

[37] The Frontier Model Forum was officially launched on July 26, 2023 (Microsoft Corporate Blogs, 2023). The companies had announced several days earlier that they would "establish or join a forum or mechanism through which they can develop, advance, and adopt shared standards and best practices for frontier AI safety" (The White House, 2023a).

1. **Frontier AI developers should adopt the NIST AI RMF, or an equivalent framework that establishes a comprehensive description of categories of defenses.** Frontier AI developers can see this first step as a "defense-in-breadth" approach that provides a precursor to defense-in-depth. Here, the value of the NIST AI RMF is not only the high-level "Govern, Map, Measure, Manage" functions, but also its breakdown of these functions into categories and subcategories, such as "Manage 4.1," "Manage 4.2," and so on. (For a full list of subcategories under the RMF, see Appendix B.)

2. **To ensure resources are well-allocated, frontier AI developers should select categories from the broader list as high-priority for a defense-in-depth approach.** Here, they can take the sub-categories listed as high priority by CLTC for catastrophic incidents from frontier AI development, such as setting risk tolerance thresholds, identifying potential catastrophic impacts, red teaming, and so on. Frontier AI developers can also borrow from other guidance, such as the forthcoming Generative AI Risk Profile that NIST is developing via its public working group.

3. **Defense-in-depth is then achieved by providing multiple independent layers for each high-priority subcategory.** For example, for the NIST AI RMF category Map 1.1, CLTC guidance suggests that organizations "Identify the potential uses, and misuses or abuses for a GPAIS, and identify reasonably foreseeable potential impacts (e.g., to fundamental rights)." Frontier AI developers could draw on risk identification techniques described by other authors such as (Koessler & Schuett, 2023), who identify techniques such as scenario analysis, using risk typologies/taxonomies, and the fishbone method. Frontier AI developers should then add further "depth" by adding measures to improve the diversity, independence, and redundancy of these techniques, such as having multiple independent teams perform this work, conducting adversarial analysis of the original analysis, and so on. As the potential risks from advanced AI systems increase, frontier AI developers should consider scaling the level of assurance (and hence costliness) of these measures appropriately.[38] Where there are insufficient measures currently available to appropriately address the level of risk, frontier AI developers may need to invest additional resources in creating such measures, and to delay deployment until adequate risk management measures can be implemented.

---

[38] For example, Anthropic has laid out a series of "AI safety levels" that map loosely to the "biosafety level" system used for dangerous pathogens, where increasingly risky AI systems require more sophisticated safety, security, and operational measures (*Anthropic's Responsible Scaling Policy, 2023*).

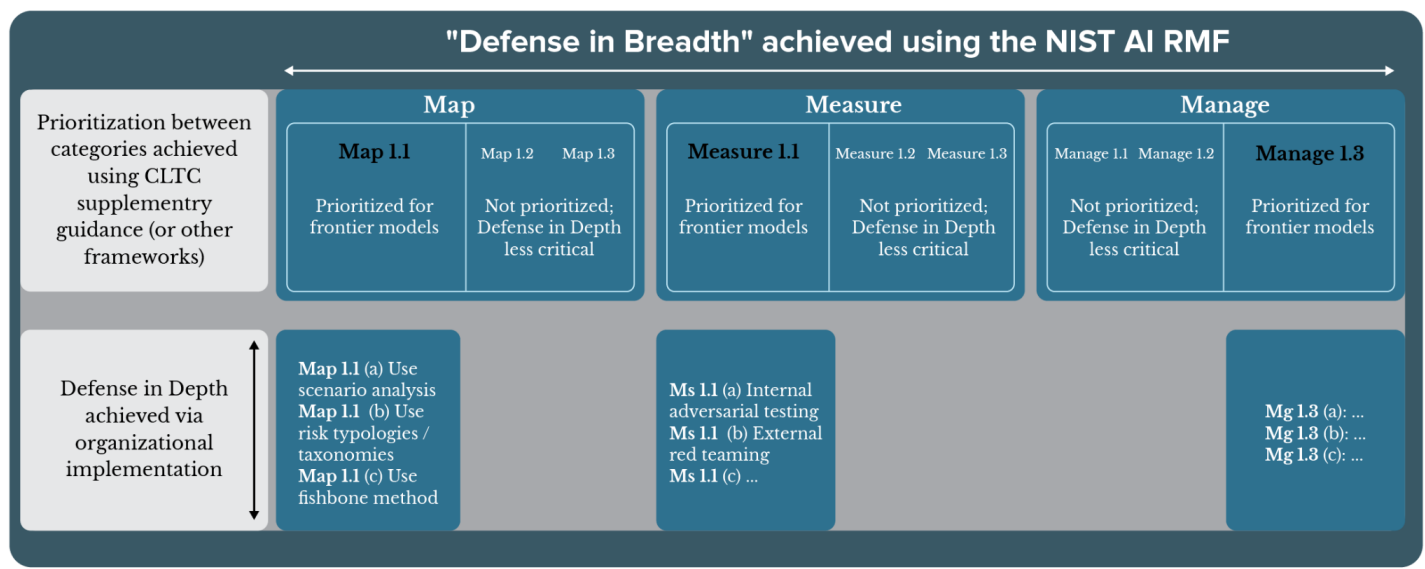

*Figure 1: Defense in Depth using the NIST AI RMF and supplementary guidance.*

## 3.4 | Limitations and future work

- **The AI RMF targets a broad audience and does not primarily address safety and security risks from frontier AI**. NIST, the Frontier Model Forum, or other researchers could adapt the AI RMF to better target these risks.

- **The AI RMF is not intended as a detailed and comprehensive list of controls to address safety and security risks from frontier AI**. A more comprehensive list of controls could facilitate frontier AI developers and policymakers adopting relevant measures, evaluating the completeness of existing measures, and comparing frontier AI developers' approaches. This would ideally be driven by NIST, but an industry body focused on AI safety and security could potentially perform a similar role (such as the [Frontier Model Forum](#)).

- **The generality of the AI RMF can make it difficult for frontier AI developers to identify gaps.** While this breadth and generality helps to highlight cross-cutting categories of activity, it can also make it more difficult to prioritize activities and assess comprehensiveness given particular use cases or threat scenarios. The threat-based framework discussed below is particularly well-suited to assuring an appropriate level of defenses against given threats, while the lifecycle model provides another perspective to evaluate comprehensiveness of measures more generally.

- **The AI RMF is not threat-specific**, and because it is focused on a developer's view, may not clearly illustrate how adversarial actors could evade or subvert existing defenses. It should therefore be paired with a threat-based approach as discussed in [Section 5](#).

# 4 | Lifecycle approach

Frontier AI developers can also adopt a lifecycle approach: describing the frontier AI development lifecycle and identifying risk management activities that the organization must perform at each phase. Lifecycle approaches in cybersecurity often emphasize the need to integrate security throughout software development, deployment, and operations, and help to promote a "safety by design" and "shift left" approach—i.e., addressing issues early in the development lifecycle, rather than waiting till the end to test and mitigate.

We suggest a six-phase framework for frontier AI development: "Plan Scope and Design Architecture; Collect and Process Data; Train and Align Model; Evaluate, Iterate, and Mitigate; Staged Deployment; and Operate and Monitor." This closely mirrors existing descriptions of the AI development lifecycle by the Organization for Economic Cooperation and Development (OECD).[39] It includes the emphasis on a "shift left" approach as mentioned above, while also calling attention to critical decision-making nodes during frontier AI deployment and the importance of monitoring and response in post-deployment system operation.

Moving forward, we suggest that the Frontier Model Forum (FMF) develop a consensus model that identifies and prioritizes key safety and security activities for developers for each of the six phases. We also suggest that frontier AI developers and research funders, including the National Science Foundation (NSF), invest in research that supports a "shift left" for frontier AI; some possible directions for research could include software requirement specification techniques borrowed from safety-critical domains, dataset curation techniques to remove potentially dangerous training data (e.g., research discussing pathogen synthesis or enhancement), and foundational research to build safer and more secure AI systems.

## 4.1 | What does this look like in cybersecurity?

Two popular frameworks for a secure software development lifecycle (SSDLC) are the Microsoft Security Development Lifecycle (SDL) and the DevSecOps framework. These are not the only widespread SSDLC frameworks, as SSDLCs vary significantly across organizations depending on their product, team needs, and threat models.[40] However, they are useful representatives of this type of model: the SDL provides a version of a SSDLC assuming a single linear process from requirements and design through to release. However, with the rise of agile methodologies, this has become dated. The newer DevSecOps framework better reflects an agile (i.e., iterative) approach to software development, but is less prescriptive and currently not well-defined.

---

[39] See Fig. 4 of Clark et al. (2022) on p. 23.
[40] There are several software development lifecycle (SDLC) frameworks, including the waterfall and agile frameworks. Each of these frameworks may in turn have multiple relevant SSDLC approaches or multiple versions thereof. Generally, the diversity of SDLCs and SSDLCs means that there is no single SSDLC framework that is as authoritative for a lifecycle-based approach to cybersecurity as the NIST CSF is for a function-based approach. See Overby (2023) for further details on SSDLCs.

One other SSDLC approach that we would particularly recommend readers review is the NIST Secure Software Development Framework (SSDF) (Souppaya et al., 2022). The NIST SSDF is organized around four groups of practices: Prepare the Organization (PO), Protect the Software (PS), Produce Well-Secured Software (PW), and Respond to Vulnerabilities (RV). We do not discuss the NIST SSDF in this section because it does not describe a linear sequential flow of events.

## 4.1.1 | Security Development Lifecycle (SDL) framework

The SDL framework focuses on five "core phases": requirements, design, implementation, verification, and release. It also includes two supporting security activities: *training* of developers, which precedes the five core phases, and *response* to incidents, which comes after the five core phases. Individual security practices are then listed in each phase. One version from 2010 that illustrates the SDL particularly clearly is provided below, listing 16 security practices.[41] The SDL was originally developed as a set of mandatory practices for internal use at Microsoft in the early 2000s, and has since gained popularity more widely.[42]

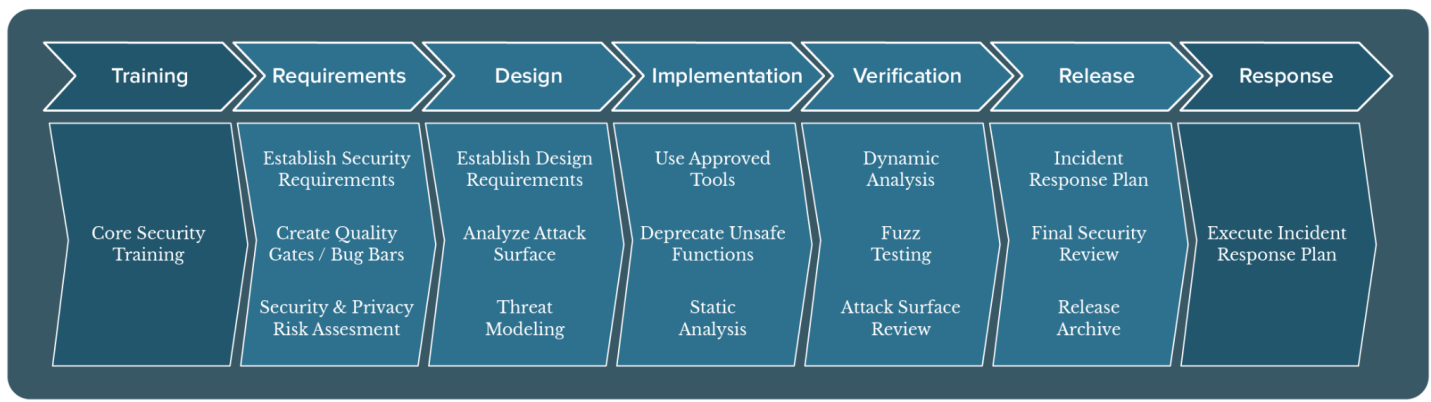

*Figure 2: Microsoft Security Development Lifecycle. Diagram content taken from "Figure 2: The Microsoft Security Development Lifecycle - Simplified" in <u>Simplified Implementation of the Microsoft SDL (2011)</u>.*

The Microsoft SDL above was originally intended to reflect a linear software development practice, but this is also why practitioners may now see the 2010 version of the SDL as being outdated.[43] This linear flow from design to implementation to testing is known as the "waterfall" model of software development, a model that has gained a reputation for being inflexible and inefficient when applied to modern software development projects that require rapid prototyping and iterative testing of new features.[44] Modern software development often follows an "agile" methodology, which focuses on shorter software development cycles (plan, design, build, test), with continuous rapid releases that incorporate customer feedback.[45]

---

[41] Microsoft has since updated these recommended security practices, reducing them from 16 to 12. Though it lists these security practices on its website, it no longer assigns them by software lifecycle phase, which is why we do not use the more current version. The current version is at *Microsoft Security Development Lifecycle Practices* (n.d.).

[42] Note that this framework is only intended to be illustrative: many of these activities do not apply to AI systems, or are not as important; for example, "static analysis" would only identify bugs in the algorithms used to train a model, not in the model itself.

[43] For example: Koussa (n.d.), which says that many methodologies like SDL "take approaches that resemble inefficient, top-down waterfall methodologies. These approaches to secure SDLC are failing many in the industry, and new approaches need to be adopted."

[44] Although the waterfall model is still valuable, especially for projects that must follow a more rigid software development process due to, e.g., safety concerns—for example, the avionics software development process, as described in standards like DO-178C (Rierson, 2013).

[45] "The Agile methodology is a project management approach that involves breaking the project into phases and emphasizes continuous collaboration and improvement. Teams follow a cycle of planning, executing, and evaluating"(Atlassian, n.d.-b). "Agile project management is an iterative approach to managing software development

## 4.1.2 | The DevSecOps framework

The DevSecOps framework better reflects this iterative approach to software development, and aims to ensure that security is addressed during both software development and IT operations (including systems administration, cloud infrastructure management, and service monitoring), particularly as these two elements become more tightly coupled.[46]

DevSecOps gained popularity in the early 2010s building on the "DevOps" framework, which had emerged a few years earlier. While DevOps focuses primarily on improving speed and efficiency by breaking down the organizational silos that often separate software development and IT operations, DevSecOps typically focuses on ensuring that security is baked into the DevOps process throughout the software development lifecycle (Alvarenga, 2022).

Both DevOps and DevSecOps rely heavily on several common elements such as cross-team collaboration, greater automation, and tight feedback loops driven by rapid prototyping, feedback, and monitoring (Alvarenga, 2022). The DevOps software development lifecycle is sometimes depicted as an infinity loop, with DevSecOps embedding security throughout the DevOps lifecycle. For example, one version is shown below:

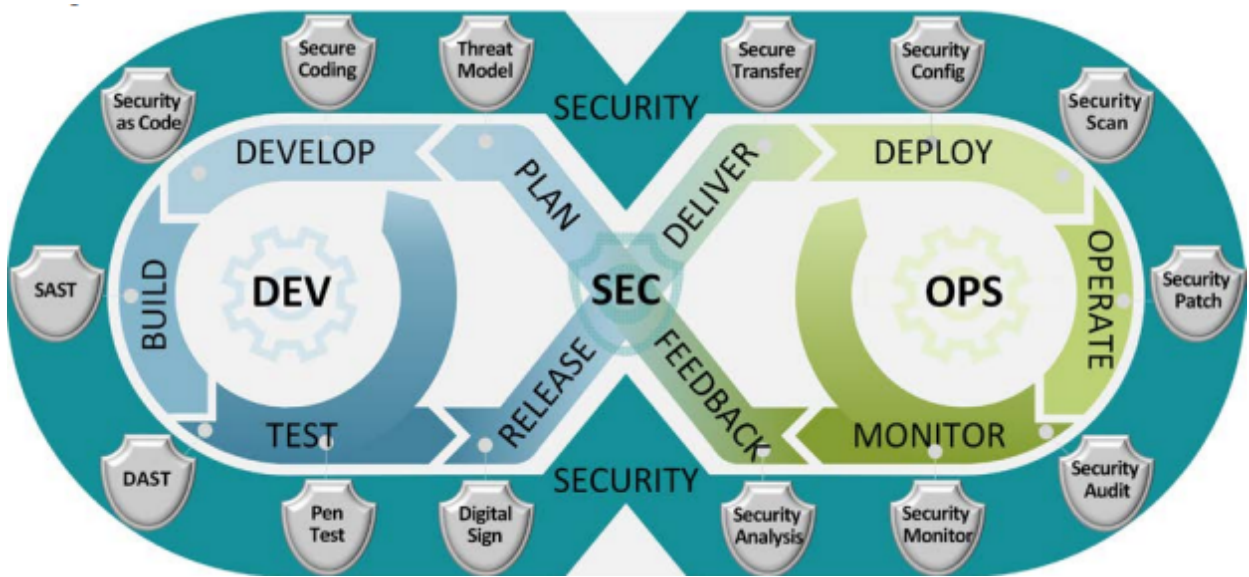

*Figure 3: DevSecOps loop. Source: Department of Defense Chief Information Officer (2019), p. 19.*

---

projects that focuses on continuous releases and incorporating customer feedback with every iteration" (Atlassian, n.d.-a).

[46] "DevSecOps helps ensure that security is addressed as part of all DevOps practices by integrating security practices and automatically generating security and compliance artifacts throughout the process" (Computer Security Division, 2020).

However, DevOps and DevSecOps are still maturing and there is still no single definitive model describing what processes are crucial to a DevSecOps approach.[47] NIST has established a project to define DevSecOps concepts and identify the key elements needed to build and document DevSecOps practices. As part of this project, NIST will draft a Special Publication to compile best practices on DevSecOps, but this Special Publication has yet to be finalized.[48]

## 4.2 | Why take a lifecycle approach?

The software lifecycle approach to security is often connected to the principle of "shifting left"–i.e., addressing security as early as possible in the lifecycle, rather than adding it on at the end.[49] Considering safety and security early in the software lifecycle can help achieve several outcomes, particularly if done in the iterative and integrated style of DevSecOps:[50]

- **Reducing the costs associated with fixing safety and security issues.** Addressing issues as they arise can make them both easier and less costly to fix, since at that point there are fewer other components entangled with or depending on them. (By comparison, imagine building a skyscraper only to discover that the steel rods in the reinforced concrete were not tested early on, and some have turned out to be defective.)

- **Mitigating safety and security risks that affect early stages of system development (e.g., model training).** For example, "data poisoning" attacks on AI systems manipulate a model's training data to change its behavior in line with an attacker's intent (Dhar, 2023).

---

[47] Almazova (2022): "Because DevOps itself is an emerging discipline with a high degree of process variations, successful DevSecOps hinges on understanding and thoughtfully integrating security into the development process."
[48] "Proposed initial activities within this DevSecOps project include: Create a new NIST Special Publication (SP) on DevSecOps practices that brings together and normalizes content from existing guidance and practices publications" (Computer Security Division, 2020).
[49] For example, the NIST Secure Software Development Framework (SSDF) says: "Most aspects of security can be addressed multiple times within an SDLC, but in general, the earlier in the SDLC that security is addressed, the less effort and cost is ultimately required to achieve the same level of security. This principle, known as shifting left, is critically important regardless of the SDLC model. Shifting left minimizes any technical debt that would require remediating early security flaws late in development or after the software is in production. Shifting left can also result in software with stronger security and resiliency" (Souppaya et al., 2022).
[50] These points are adapted from NIST's description of the value of DevSecOps (Computer Security Division, 2020):

- **"Reduces vulnerabilities, malicious code, and other security issues** in released software without slowing down code production and releases
- **Mitigates the potential impact of vulnerability exploitation throughout the application lifecycle**, including when the code is being developed and when the software is executing on dynamic hosting platforms
- **Addresses the root causes of vulnerabilities to prevent recurrences**, such as strengthening test tools and methodologies in the toolchain, and improving practices for developing code and operating hosting platforms
- **Reduces friction between the development, operation, and security teams** in order to maintain the speed and agility needed to support the organization's mission while taking advantage of modern and innovative technology"

We have modified these points for generality so that they also can be applied to AI systems.

- **Addressing the root causes of safety and security issues to prevent recurrences.** For example, flaws in the development toolchain–such as inadequate tools and methods for software testing–may lead to organizations persistently missing certain classes of issues.

- **Reducing friction between the development, operation, and safety/security teams.** Ensuring that developers are adhering to secure coding practices can be easier than fixing their bugs afterwards.[51] Some operational and safety/security tasks also require developer assistance: for example, it is best to consider having a good logging and monitoring pipeline for security alerts early in the development process. If logging is only considered later in the development process, code may be implemented in a way that makes key security events difficult to log or interpret.

Mapping the model lifecycle of frontier AI development also has additional benefits. Firstly, by separating stages, it draws attention to critical decision-making nodes in the deployment process, such as decisions on whether to deploy and/or to open-source frontier models. Secondly, it highlights that safety activities should not stop after deployment, via inclusion of post-deployment stages. This draws attention to the fact that incidents may arise during operations or subsequent updates of the model.[52] The need for organizations to address incidents during operations is well-recognized in cybersecurity, including under the NIST CSF and the DevSecOps framework, but practices for this are still underdeveloped in AI safety and security.[53]

## 4.3 | Usage for frontier AI governance

Below, we summarize general descriptions of the AI development lifecycle by the OECD and NIST and details of frontier model development as described by OpenAI, Microsoft, and other developers. These then inform a possible framework that we propose for a lifecycle-based approach to governing frontier AI systems.

### 4.3.1 | Existing descriptions of the AI development lifecycle

The Organization for Economic Cooperation and Development (OECD) has drafted a general model for the AI development lifecycle.[54] The NIST AI RMF also adopts this model and includes it as part of the explanatory preface to the function-based RMF.[55] The NIST description of the model includes six broad phases:[56]

---

[51] For some tasks that require unusual degrees of expertise or security-related context, it may be easier for security experts to do things themselves (during development or in testing) rather than imparting best practices to the software engineers. However, we expect it to usually be the case that it is easier for security experts to teach best practices to other engineers.

[52] For example, the GPT-4 system card says: "Be cognizant of, and plan for, capability jumps "in the wild": Methods like fine-tuning and chain-of-thought prompting could lead to capability jumps in the same base model. This should be accounted for explicitly in internal safety testing procedures and evaluations" (OpenAI, 2023b, p. 69).

[53] E.g., see O'Brien et al. (2023).

[54] See Fig. 4 of Clark et al. (2022) on p. 23.

- **Plan and Design:** Articulate and document the system's concept and objectives, underlying assumptions, and context in light of legal and regulatory requirements and ethical considerations.
- **Collect and Process Data:** Gather, validate, and clean data, and document the metadata and characteristics of the dataset, in light of objectives, legal and ethical considerations.
- **Build and Use Model:** Create or select algorithms; train models.
- **Verify and Validate:** Verify & validate, calibrate, and interpret model output.
- **Deploy and Use:** Pilot, check compatibility with legacy systems, verify regulatory compliance, manage organizational change, and evaluate user experience.
- **Operate and Monitor:** Operate the AI system and continuously assess its recommendations and impacts (both intended and unintended) in light of objectives, legal and regulatory requirements, and ethical considerations.

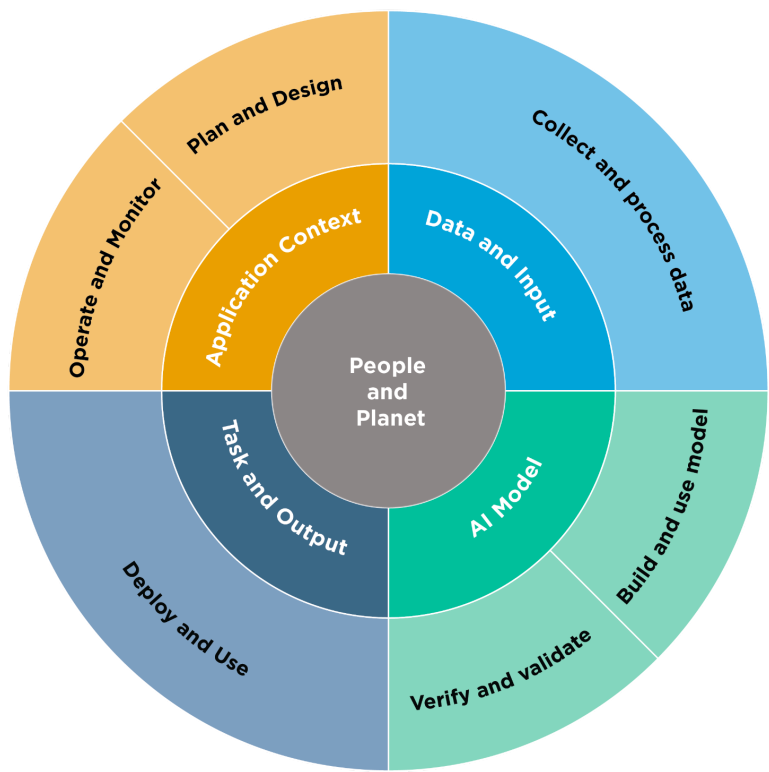

*Figure 4: OECD/NIST AI development lifecycle model. Source: (Tabassi, 2023).*

---

[55] The graphic in this report is taken from Fig. 2 (p. 10) of the NIST AI RMF (Tabassi, 2023).
[56] Content taken from Fig. 3 (p. 11) of the NIST AI RMF (Tabassi, 2023). The NIST AI RMF includes more details including on the specific actors who should be involved in each phase; see Appendix A of the AI RMF. This graphic includes a seventh category, "Use or Impacted By: Use system/technology; monitor & assess impacts; seek mitigation of impacts, advocate for rights," which we omit because it is not included in Fig. 2, and due to it being most relevant to downstream users and deployers, rather than developers.

As with the NIST AI RMF at large, this lifecycle model is meant to be generally applicable, and may differ from how specific frontier AI models are developed in practice. We can compare the OECD/NIST lifecycle framework with OpenAI's own description of its development and deployment lifecycle for LLMs, as below. (This lifecycle description may not be generalizable to models trained by other labs, or in other domains.)

OpenAI's description of its development and deployment lifecycle includes five main stages ([Brundage et al., 2022](#)):

- **Initial Development:** problem identification and goal setting; initial impact assessment; data sourcing, curation, and filtration
- **Alignment:** instruction generation; fine-tuning; alignment evaluations
- **Evaluation and Iterative Development:** model evaluations; revised impact assessment and hazard analysis; red teaming and user testing
- **Deployment and Ongoing Evaluation:** private betas; use case pilots; misuse detection and response
- **Downstream Assessment:** retrospective reviews; retrospective impact assessment; platform-level risk measurement

The most notable difference between OpenAI's and the OECD's model is that the OpenAI model refers to an "alignment" phase. Currently, the model training that OpenAI conducts can be viewed loosely as two distinct phases: first, OpenAI trains a "base model" on a large amount of data; subsequently, OpenAI finetunes this base model to remove harmful behavior, using a technique known as "reinforcement learning with human feedback" ([Lowe & Leike, 2022](#); [OpenAI, 2023b](#)). Other companies may use different alignment techniques; for example, Anthropic uses an approach called "constitutional AI," relying on feedback from an AI system instead of human feedback ([Bai et al., 2022](#)). Because approaches to alignment are still changing, we frame this overall process as a single "Train and Align Model" phase in our lifecycle description below.

Other researchers have also laid out alternative lifecycle frameworks, which may be useful for future framework developers to tap on.[57]

## 4.3.2 | Proposed lifecycle framework

The below framework draws closely from the OECD and NIST frameworks, while integrating additional details from frontier models. Frontier models have large compute requirements and high training costs, making their development cycles lengthier than many smaller AI projects. Because many frontier models are general-purpose and can be used in an extremely wide range of contexts, they also require an extensive pre-deployment testing regimen and a phased approach to rollout.

---

[57] E.g., [Fang et al. (2023)](#) outlines eight high-level phases: model requirement, data collection, data preparation, feature engineering, model training, model evaluation, system development, and model monitoring. [De Silva & Alahakoon (2022)](#) lays out a 19-stage lifecycle. There are also alternatives laid out in textbooks, e.g., [Thomas et al. (2021)](#).

- **Plan Scope and Design Architecture:** Identify the scope, objectives, and expected capabilities of the model, and document them. Specify data and compute requirements and model architecture, including safety and alignment features incorporated into the system design. Conduct initial impact assessment and identify and prioritize potential affected stakeholders. If possible, specify behavioral requirements for subsequent go/no-go decisions (e.g., safety and security requirements), and have external parties with relevant expertise rigorously review organizational safety plans.

- **Collect and Process Data:** "Gather, validate, and clean data and document the metadata and characteristics of the dataset" ([Tabassi, 2023](#)). Filter data to remove inappropriate training data, e.g., violent content, using human review and/or automated tools.[58] Verify the integrity of data against malicious attacks, e.g., data poisoning.

- **Train and Align Model:** Train the base model on a secure compute cluster, limiting access to base model weights to minimize risk of theft or misuse. Finetune the model using alignment methods, e.g., reinforcement learning with human feedback (RLHF), or constitutional AI.[59]

- **Evaluate, Iterate, and Mitigate:** Conduct internal and external testing to assess safety and security of the model, including red teaming to anticipate potentially harmful behavior by deliberately eliciting such behavior in a safe environment. Mitigate harms via finetuning of model weights or implementation of other guardrails. If necessary, notify and coordinate with other actors to delay model development and/or develop countermeasures if harms are serious and cannot be adequately mitigated. (This phase could overlap with "Train and Align Model.")

- **Staged Deployment:** Release model to trusted parties in stages (e.g., through private betas or use case pilots). Before significant deployments, conduct pre-deployment risk assessment(s) and be willing to not release the model if doing so is assessed to be too high-risk. Decide what information is not safe to publicly release (e.g., training details, model weights) based on potential risk of abuse by malicious actors.

- **Operate and Monitor:** "Operate the AI system and continuously assess its recommendations and impacts (both intended and unintended)" ([Tabassi, 2023](#)). Monitor for anomalies, misuse, and systemic societal effects and respond as appropriate, including limiting access to the model if needed ([O'Brien et al., 2023](#)). Continue to finetune the model to improve safety and security based on observed real-world behavior.

---

[58] E.g., for GPT-4, OpenAI implemented dataset interventions: "At the pre-training stage, we filtered our dataset mix for GPT-4 to specifically reduce the quantity of inappropriate erotic text content. We did this via a combination of internally trained classifiers and a lexicon-based approach to identify documents that were flagged as having a high likelihood of containing inappropriate erotic content. We then removed these documents from the pre-training set" ([OpenAI, 2023b, p. 61](#)).

[59] On RLHF, see [Christiano et al. (2017)](#). On constitutional AI, see [Bai et al. (2022)](#).

## 4.3.3 | Discussion of proposed framework

In line with the "shift left" approach in DevSecOps, the above life cycle suggests that it may be worth trying to push for a similar "shift left" in AI safety and security. In the policy debate on AI safety and security, current proposed interventions skew heavily toward the latter part of the development cycle: i.e., testing, evaluation, and developing mitigations for any issues that are discovered.[60] As models become increasingly powerful and complex, scaling this test-and-mitigate approach will become more challenging, making it important to address issues as early in the development cycle as possible.

A "safety by design" or "security by design" approach for frontier models could incorporate the following measures:

- **Plan and Design: Organizations could examine software requirement specification techniques in safety-critical domains, e.g., autonomous vehicles, and extend them to frontier models.** In high-reliability software engineering, one of the methods used in safety-critical software is specifying detailed "requirements"–i.e., descriptions of *what* the software should do, rather than just *how* the software should do it ([Rierson, 2013](#)). Effective software requirements can ensure that necessary safety requirements are implemented and that the system has no unwanted functionality that could contribute to an accident.[61]

  Adapting these techniques for frontier models would require substantial innovation, but could pay off. It would be extremely challenging to meet the demanding criteria conventionally used in aviation, for example, because these techniques depend largely on code being interpretable by humans. However, other safety-critical disciplines adopting AI–such as autonomous vehicles–have grappled with how to develop behavioral requirements for safety, and this work could perhaps be extended to frontier AI models.[62]

  Increased safety during the planning and design phase could also include investing in research and development to develop architectures that inherently support alignment and/or safety, or restricting use cases of frontier AI models upfront in order to make the risks more predictable and manageable.

- **Collect and Process Data: Organizations could invest in and share information on dataset curation techniques to remove training data that may contribute to harmful outputs.** Such techniques are already in use; for example, OpenAI has documented its use of human-machine teams to remove graphic and explicit images from the training set for DALL-E, and to remove erotic content from the training set for GPT-4 ([Nichol, 2022](); [OpenAI, 2023b, p. 61]()). Eliminating harmful data from models pre-training could help to mitigate the model later producing harmful output, because it is difficult to reliably prevent a model from reproducing harmful output once it has learned that output from a given dataset.[63] Selectively removing potentially harmful information from the training dataset, such as research on the creation or enhancement of pathogens, could potentially reduce malicious users' ease of access to such data ([Soice et al., 2023]()).

---

[60] For example, see our discussion of the voluntary commitments from leading AI labs in [Section 6.2](#).
[61] More specifically, this borrows from the idea of "bidirectional traceability" as articulated by the standards document DO-178C for safety-critical software. DO-178C requires bidirectional traceability for the most safety-critical level of software (DAL A), i.e., showing both that all necessary safety requirements are implemented in code ("forward traceability"), and that there is no "dead code" that is not described by a requirement and could cause an accident through unwanted functionality ("backward traceability") ([Rierson, 2013](#)).
[62] For example, [Madala et al. (2023)](#) and [Q. A. D. S. Ribeiro et al. (2022)](#) discuss challenges associated with requirements engineering and requirements specification for autonomous vehicles.

- **Train and Align Model: Organizations could invest in foundational research to build safer and more secure AI systems.** For example, one line of research by OpenAI involves developing tools and techniques for "scalable oversight," which involves using AI systems to facilitate evaluation of other AI systems. OpenAI has stated that it will allot 20% of its computing resources to pursuing this and related goals ([Leike & Sutskever, 2023]()).[64] Other frontier AI labs have also created formal "safety teams" or "alignment teams," such as DeepMind, Anthropic, and Inflection.

Policymakers should also consider funding technical work to improve safety and security on the "left" side of the frontier model development lifecycle (i.e., planning, data collection, and training methods). Given that some researchers have raised fundamental concerns about the safety and security of current AI technologies,[65] a timely injection of funding could help discover new approaches that might avert these flaws from being magnified as systems become more capable and integrated into society. By comparison with cybersecurity, leading computer scientist Tony Hoare has called his invention of the null pointer in the 1960s a "billion dollar mistake" due to its frequent exploitation by malicious actors to conduct cyberattacks in subsequent decades. ([Hoare, 2009]()).

The need to develop AI that is "safe by design" falls under one of nine main strategic thrusts described in the 2023 update to the National AI Research and Development (R&D) Strategic Plan.[66] Given this, the National Science Foundation (NSF) should lead investment in a "shift left" on AI safety and security, continuing to expand on its existing funding for trustworthy AI.[67] Other departments and agencies should also invest appropriately in such efforts, particularly the Department of Defense (DoD), the Department of Energy (DOE), and the Department of Health and Human Services (HHS), which are the other government bodies besides the NSF that funded more than $200 million dollars each of AI R&D in FY 2022 (National Artificial Intelligence Research Resource Task Force, 2023, p. 10). Ensuring adequate investment in R&D for safety by design will become especially important as departments and agencies step up overall AI R&D investment; the DOE, for example, has proposed a potential initiative to Congress that could entail billions of dollars in funding (Taylor, 2023).

---

[63] "The challenge lies in the fact that, once learned, it is virtually impossible to 'remove' knowledge from these models—the information remains embedded in their neural networks. This means safety mechanisms primarily work by preventing the model from revealing certain types of information, rather than eradicating the knowledge altogether" ([Volodin & Vanunu, 2023]()). Some researchers have been investigating a set of techniques known as "machine unlearning," which does aim to remove knowledge from models ([Duffin, 2023]()). However, curating data is not foolproof as models may still be able to generalize from existing datasets to produce harmful outputs even if they are not explicitly trained on examples of such output.

[64] This can be seen as somewhat analogous to how DevSecOps equips developers with security tools that they can use themselves to eliminate vulnerabilities.

[65] For example, researchers have demonstrated that all major large language models are vulnerable to a common series of attacks that can be generated automatically, which could make it impossible to effectively secure them (*Universal and Transferable Attacks on Aligned Language Models,* n.d.; [Zou et al., 2023]()).

[66] See "Strategy 4: Ensure the Safety and Security of AI Systems" at pp. 16-17 of [Select Committee on Artificial Intelligence & National Science and Technology Council (2023)](): "The process of securing and making safe AI… must be incorporated in all stages of the AI system life cycle, from the initial design and data/model building to verification and validation, deployment, operation, and monitoring. "Safety by Design" must therefore be an important part of the AI R&D portfolio."

Ideally, policymakers would also use regulation to encourage a "safe by design" approach to AI systems; see for example the proposal in (Anderljung et al., 2023). However, such an approach is out of scope for this report.

## Deployment and post-deployment measures

Following deployment, frontier AI developers must also be prepared to implement "deployment corrections" to address potential or observed dangerous behavior, use, or outcomes from deployed models (O'Brien et al., 2023). Even if frontier AI developers implement strong pre-deployment risk assessments, these risk assessments may not be able to identify all risks, and future performance improvements, integrations, interactions, or uses of the model may introduce new risks.

Such deployment corrections can include user-based restrictions, access frequency limits, capability or feature restrictions, use case restrictions, or model shutdown. These measures exist on a spectrum and can be combined as appropriate. They must also be supported by other capabilities, such as setting thresholds for implementing deployment corrections, and continuous logging and monitoring. Below, we provide a diagram illustrating the measures that frontier AI developers may need to implement to ensure robust post-deployment incident response.

---

[67] One report by the Federation of American Scientists (FAS) estimates that 10-15% of annual AI funding from the NSF's Computer and Information Science and Engineering (CISE) directorate goes toward "trustworthy AI," which includes efforts that support interpretability/explainability, fairness/non-discrimination, robustness/safety, and privacy preservation (Alexander & Kaushik, 2023). Given that trustworthy AI also includes multiple strategic thrusts under the National AI R&D Strategic Plan–e.g., Strategy 3 on the ethical, legal, and societal implications of AI, or Strategy 6 on measuring and evaluating AI systems–there is arguably room to expand this funding beyond 10-15% as AI becomes increasingly important in daily life.

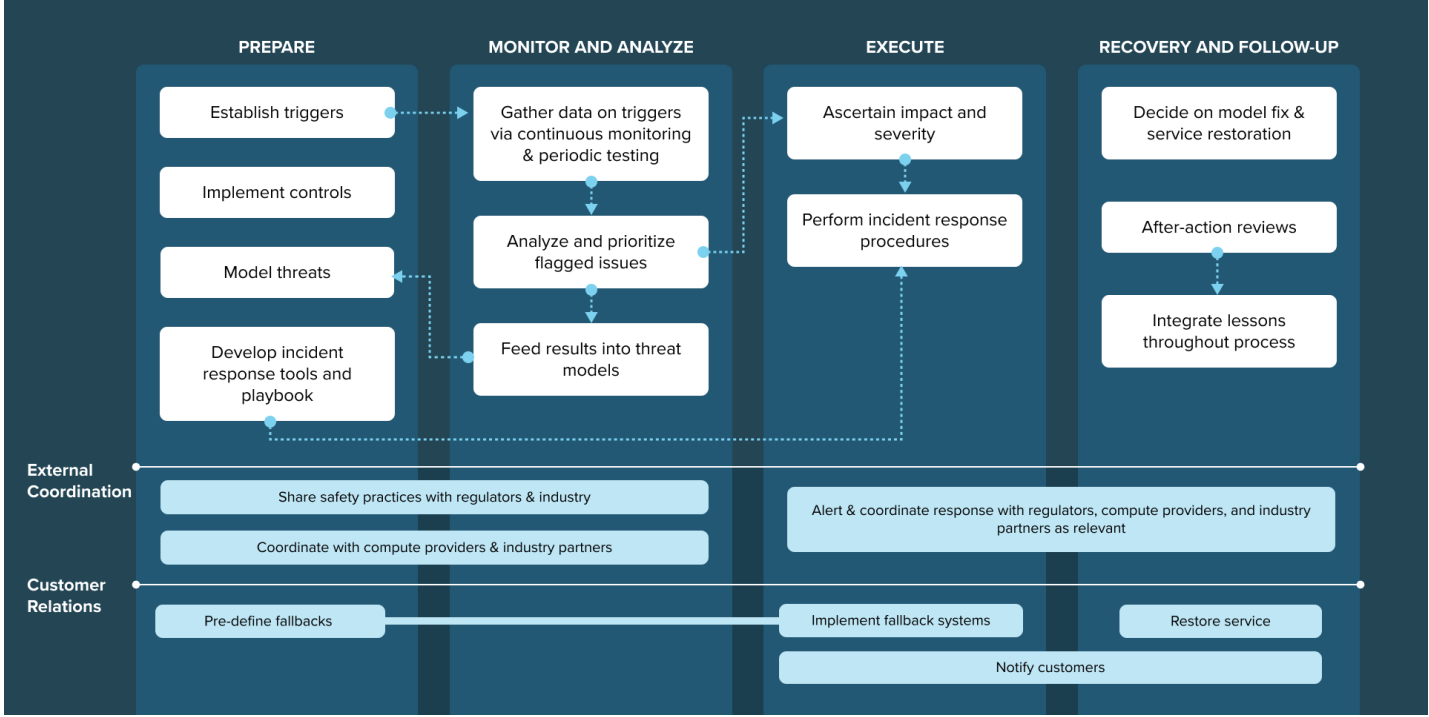

*Figure 5. An end-to-end process for implementing deployment corrections for frontier AI models. Source: O'Brien et al. (2023).*

Further details are available in our report, "Deployment corrections: an incident response framework for frontier AI models" (O'Brien et al. 2023).

## 4.4 | Limitations and future work

- **The lifecycle approach may not effectively capture functions and measures that cut across multiple phases.** As a result, it may be difficult to assess if there are any gaps in the organization's implementation of these functions and measures using the lifecycle approach alone.

    For example, the NIST AI RMF's "Govern" function, which describes overarching measures that enable other functions, could exist outside of or parallel to the lifecycle model as it involves measures such as assessing the adequacy of other implemented measures.[68] Other functions like transparency and disclosure on model capabilities and limitations, or cybersecurity and insider security, may similarly cut across multiple phases.[69] To ensure adequate coverage, organizations

---

[68] E.g., see Govern 1.5: "Ongoing monitoring and periodic review of the risk management process and its outcomes are planned, organizational roles and responsibilities are clearly defined, including determining the frequency of periodic review." A more detailed overview of the NIST AI RMF Govern function is available at Appendix B.
[69] For example, Fig. 4 of Shevlane et al. (2023) describes transparency and security as cutting across multiple parts of the lifecycle—before training, during training, pre-deployment, and post-deployment.

should use complementary approaches like the functional approach in tandem with the lifecycle approach.

- **The lifecycle approach currently lacks the granularity of the functional approach and is less ready to implement.** Currently, there is no commonly agreed upon version for a frontier AI lifecycle model that, for example, subdivides phases into categories or associates them with individual activities. This makes it more difficult to assess whether defenses are truly comprehensive. However, this could be mitigated by the development of a detailed consensus lifecycle model, e.g. by the Frontier Model Forum.

- **The lifecycle model centers on the organization's own activities, which could lead organizations to neglect interactions with other systems and organizations as being "out of frame."** Other actors may spend substantial effort trying to stretch the capabilities of the frontier model or build other architecture on top of the model; as an example, following the launch of GPT-4, an extensive developer community sprung up using GPT-4 to build autonomous agents (*Insight: Race towards "autonomous" AI Agents Grips Silicon Valley | Reuters*, n.d.). Organizations should be explicit about needing to anticipate, monitor, and address events out of their immediate control, particularly in the latter stages of the model lifecycle.

# 5 | Threat-based approach

Frontier AI developers can also take an adversary's perspective to verify if their defenses are adequate against malicious or subversive activity from non-state actors, nation-states, or sophisticated goal-directed AI systems. Such threat-based approaches are stronger when they tap on knowledge bases of adversary behavior, such as the MITRE ATT&CK database, which compiles empirical data and research from cybersecurity to describe the tactics, techniques, and procedures (TTPs) that malicious actors use. Organizations can use such databases to ensure that they have adequate defensive coverage, such as by simulating attacks from adversaries.

Unlike the functional and lifecycle approaches, threat-based approaches are not effective in scenarios where harm is not caused by specific malicious actors—for example, structural issues like misinformation or mass unemployment. However, the threat-based lens is valuable because defenses that seem reasonable from a developer's perspective may not be sufficient when facing sophisticated goal-directed actors, who actively search for and optimize against defensive gaps.

Moving forward, we suggest that MITRE and/or the Frontier Model Forum establish and build out knowledge bases to facilitate threat-based approaches. The MITRE ATLAS database is a version of ATT&CK focused on machine learning, which MITRE and partners should continue to develop. We suggest two potential changes for ATLAS: (1) add a version of ATLAS that uses categories familiar to machine learning experts, in addition to the current version that uses categories familiar to cybersecurity practitioners; (2) expand ATLAS to cover TTPs that use AI systems to affect other systems ("effect on world"), in addition to TTPs to attack systems ("effect on model"). We also recommend the US Cybersecurity and Infrastructure Agency (CISA) assess the effects of frontier AI systems on the top ten most vulnerable National Critical Functions.

## 5.1 | What does this look like in cybersecurity?

The MITRE Adversarial Tactics, Techniques and Common Knowledge (MITRE ATT&CK) framework is a knowledge base that documents cybersecurity tactics and techniques used by malicious actors (*MITRE ATT&CK®, n.d.-b*). As of 2023, the framework lists over 300 individual techniques (e.g., "Steal or Forge Kerberos Tickets") that are organized into 14 tactics (e.g., "Reconnaissance," "Defense Evasion," "Command and Control"). While the 14 tactics represent the "why" of a technique–i.e., the high-level objective an adversary wants to accomplish–the 300 techniques represent "how" the adversary achieves this objective, and/or "what" specifically they gain (Strom, 2020).

MITRE ATT&CK is probably the single threat-based framework that is most commonly used by cybersecurity practitioners. In a 2020 survey of security professionals from 325 large- and medium-sized enterprises in the UK, US, and Australia, 81% of enterprises reported using ATT&CK.[70] The primary use of ATT&CK is helping defenders to understand the tactics that adversaries may use against them, identifying gaps in monitoring and security tools, and then adjusting their defenses appropriately.[71] 57% of respondents to the 2020 survey said that ATT&CK was "helpful for determining gaps in currently deployed security tools" (Basra & Kaushik, 2020, p. 2).[72]

Because MITRE ATT&CK provides a high level of detail, including observations about what tactics specific threat actors use, defenders can use it to simulate attacks by the threat actors they are most worried about. For example, US government agencies could refer to the techniques that are commonly used by APT28 and APT29, two Russia-linked groups that have been linked to the 2020 SolarWinds incident and espionage campaigns in 2015-16 against the Hillary Clinton campaign and the Democratic National Committee (Naeem et al., 2023, p. 29). The ATT&CK database is updated biannually based on publicly available information, which is relatively frequent compared to other standards (e.g., the NIST CSF, which is only refreshed every several years).[73]

---

[70] "Eighty-one percent of enterprises in our survey currently use the ATT&CK framework in general" (Basra & Kaushik, 2020, p. 11). These are similar statistics to those reported in Oltsik (2022): "According to ESG research, 48% of organizations say they use the MITRE ATT&CK framework "extensively" for security operations while another 41% use it on a limited basis."

[71] For example, MITRE itself provides four common use cases for MITRE ATT&CK: detection and analytics; threat intelligence; adversary emulation and red teaming; and assessment and engineering (*Getting Started | MITRE ATT&CK®, n.d.*).

[72] This is lower than one might hope for (as it implies that 43% of respondents did *not* see ATT&CK as useful for this purpose)–see 5.4 | Limitations and future work as for some discussion of why.

[73] See MITRE ATT&CK® (n.d.-a) on where this information is taken from: "Publicly available threat intelligence and incident reporting is the main source of data in ATT&CK. We take what's available in the public and distill out common TTPs. We also use publicly available research on new techniques that closely align with what adversaries commonly do since new TTPs often get used in the wild quickly."

| Initial Access | Execution | Persistence | Privilege Escalation | Defense Evasion | Credential Access | Discovery |
|---|---|---|---|---|---|---|
| 9 techniques | 14 techniques | 19 techniques | 13 techniques | 42 techniques | 17 techniques | 31 techniques |
| Drive-by Compromise | Command Administration Command | Account Manipulation (5) | Abuse Elevation Control Mechanism (4) | Abuse Elevation Control Mechanism (4) | Adversary-in-the-Middle (3) | Account Discovery |
| Exploit Public-Facing Application | Command and Scripting Interpreter (9) | BITS Jobs | Access Token Manipulation (5) | Access Token Manipulation (5) | Brute Force (4) | Application Window Discovery |
| External Remote Services | Container Administration Command | Boot or Logon Autostart Execution (14) | Boot or Logon Autostart Execution (14) | BITS Jobs | Credentials from Password Stores (5) | Browser Information Discovery |
| Hardware Additions | Deploy Container | Boot or Logon Initialization Scripts (5) | Boot or Logon Initialization Scripts (5) | Build Image on Host | Exploitation for Credential Access | Cloud Infrastructure Discovery |
| Phishing (3) | Exploitation for Client Execution | Browser Extensions | Create or Modify System Process (4) | Debugger Evasion | Forced Authentication | Cloud Service Dashboard |
| Replication Through Removable Media | Inter-Process Communication (3) | Compromise Client Software Binary | Domain Policy Modification (2) | Deobfuscate/Decode Files or Information | Forge Web Credentials (2) | Cloud Service Discovery |
| Supply Chain Compromise (3) | Native API | Create Account (3) | Escape to Host | Deploy Container | Input Capture (4) | Cloud Storage Object Discovery |
| Trusted Relationship | Scheduled Task/Job (5) | Create or Modify System Process (4) | Event Triggered Execution (16) | Direct Volume Access | Modify Authentication Process (8) | Container and Resource Discovery |
| Valid Accounts (4) | Serverless Execution | Event Triggered Execution (16) | Exploitation for Privilege Escalation | Domain Policy Modification (2) | Multi-Factor Authentication Interception | Debugger Evasion |
| | Shared Modules | External Remote Services | Hijack Execution Flow (12) | Execution Guardrails (1) | Multi-Factor Authentication Request Generation | Device Driver Discovery |
| | Software Deployment Tools | Hijack Execution Flow (12) | Process Injection (12) | Exploitation for Defense Evasion | Network Sniffing | Domain Trust Discovery |
| | System Services (2) | Implant Internal Image | Scheduled Task/Job (5) | File and Directory Permissions Modification (2) | OS Credential Dumping (8) | File and Directory Discovery |
| | User Execution (3) | Modify Authentication Process (8) | Valid Accounts (4) | Hide Artifacts (10) | Steal Application Access Token | Group Policy Discovery |
| | Windows Management Instrumentation | Office Application Startup (6) | | Hijack Execution Flow (12) | Steal or Forge Authentication Certificates | Network Service Discovery |
| | | Pre-OS Boot (5) | | Impair Defenses (10) | Steal or Forge Kerberos Tickets (4) | Network Share Discovery |
| | | Scheduled Task/Job (5) | | Indicator Removal (9) | | Network Sniffing |
| | | | | Indirect Command Execution | | Password Policy Discovery |
| | | | | Masquerading (8) | | Peripheral Device Discovery |
| | | | | Modify Authentication Process (8) | | Permission Groups Discovery (3) |
| | | | | Modify Cloud Compute Infrastructure (4) | | |
| | | | | Modify Registry | | |

Figure 6: Excerpt from MITRE ATT&CK. Source: attack.mitre.org (taken on Aug 5, 2023); full MITRE ATT&CK database is not shown due to size. © 2023 The MITRE Corporation. This work is reproduced and distributed with the permission of The MITRE Corporation.

### 5.1.1 | An alternative threat-based approach: the kill chain

MITRE ATT&CK is not the only threat-based approach that companies can use to assess their defenses. It contrasts with the concept of a "kill chain"–a sequential set of steps that describes the high-level objectives in an attacker campaign. The most popular kill chain is the Lockheed Martin Cyber Kill Chain, which runs through seven steps: Reconnaissance, Weaponization, Delivery, Exploitation, Installation, Command & Control, and Actions on Objectives (Lockheed Martin, n.d.).

In theory, by intervening at any stage in the attack, a defender can break the chain and disrupt the entire attack.[74] By contrast, MITRE ATT&CK explicitly is *not* a kill chain; the 14 high-level tactics are not sequential, and the framework assumes that attackers can skip stages or jump back and forth between them.[75]

We focus on ATT&CK, not the Cyber Kill Chain, because the "kill chain" is an idealized description of an attack campaign that, in practice, attackers do not always follow. Moreover, because it describes adversaries at a high level, linking concrete defensive measures to the Cyber Kill Chain can be difficult—this was one of the motivations driving the creation of ATT&CK.[76] Given the diversity of safety and security risks inherent to frontier AI systems, a "kill chain" approach to threat modeling could be too simplistic to capture important information and design useful defenses.

## 5.2 | Why take a threat-based approach?

A threat-based approach to frontier AI development emphasizes an adversary's perspective, setting it apart from the functional and lifecycle approaches, which emphasize the developer's (or defender's) perspective. By thinking like an attacker, defenders can ensure that the measures they have implemented are adequate against the types of attacks they expect to face. As MITRE describes it, ATT&CK allows defenders "to follow the adversary's motivation for individual actions and understand how the actions and dependencies relate to specific classes of defenses that may be deployed in an environment."[77]

This approach is valuable because defenses that seem reasonable from a developer's perspective may not be sufficient when facing sophisticated goal-directed actors. Such actors can be highly motivated to find creative failure modes for AI systems by searching for gaps in developers' defensive postures and optimizing against them. Examples of such skilled, persistent actors could include actors such as nation-states, criminal groups, or other non-state actors. Potentially, they could also include sophisticated AI systems that are designed to pursue long-term goals strategically with high levels of autonomy (Brown, 2023). However, the focus on adversarial action also makes the threat-based approach less useful for mitigating risks that do not involve a specific malicious actor: for example, structural harms such as the propagation of misinformation or mass unemployment.

---

[74] From *Gaining the Advantage: Applying Cyber Kill Chain® Methodology to Network Defense* (2015), p. 3: "Stopping adversaries at any stage breaks the chain of attack! Adversaries must completely progress through all phases for success; this puts the odds in our favor as we only need to block them at any given one for success."
[75] Addressing the question "What is the relationship between ATT&CK and the Lockheed Martin Cyber Kill Chain®?", the ATT&CK FAQ says: "ATT&CK and the Cyber Kill Chain are complementary. ATT&CK sits at a lower level of definition to describe adversary behavior than the Cyber Kill Chain. ATT&CK Tactics are unordered and may not all occur in a single intrusion because adversary tactical goals change throughout an operation, whereas the Cyber Kill Chain uses ordered phases to describe high level adversary objectives" (MITRE ATT&CK®, n.d.-a).
[76] Strom (2020) describes one of four motivations for creating MITRE ATT&CK as: "Lifecycle models that didn't fit. Existing adversary lifecycle and Cyber Kill Chain concepts were too high-level to relate behaviors to defenses — the level of abstraction wasn't useful to map TTPs to new types of sensors."
[77] See Section 4.1.1, "Adversary's Perspective," of Strom et al. (2018), pp. 20-21. More generally, MITRE notes that ATT&CK was created with three conceptual ideas in mind: (1) maintaining the adversary's perspective, (2) following real-world use of activity through empirical use examples, and (3) operating at a level of abstraction appropriate to bridge offensive action with possible defensive countermeasures.

Conceptually, MITRE ATT&CK was created as a "mid-level adversary model," describing adversary behaviors in enough detail to map them to defenses, while maintaining enough abstraction to be generalizable across many threats and systems (Strom et al., 2018, pp. 22-23). In this sense, it sits in between high-level models like the Lockheed Martin Cyber Kill Chain (described above), and low-level databases of exploits and vulnerabilities. MITRE originally created ATT&CK to facilitate adversary emulation exercises, so that teams could assess how comprehensive their defensive coverage was against simulated attacks.[78]

ATT&CK is now used across a variety of cases, including adversary emulation (using threat intelligence to simulate an adversary), red teaming ("applying an adversarial mindset without… threat intelligence"), detecting potential malicious behavior on a system or network, assessing defensive gaps, assessing organizational maturity, or understanding the TTPs of prominent adversarial groups (Strom et al., 2018, p. 3). While threat-based approaches for defending against subversion and misuse of frontier AI systems are still under development, we expect that establishing common knowledge bases could facilitate some or all of these use cases for frontier AI systems, particularly defensive gap assessment.

## 5.3 | Usage for frontier AI governance

### 5.3.1 | Existing work

The MITRE Adversarial Threat Landscape for Artificial-Intelligence Systems (MITRE ATLAS) database is the counterpart of MITRE ATT&CK for machine learning (ML) systems. As its name suggests, it focuses primarily on adversarial tactics, such as tactics to induce misclassifications, missed detections, or unauthorized model output, and does not cover other issues such as unsafe behavior or misuse scenarios.

ATLAS currently adopts the same high-level categories ("tactics") for ML systems as ATT&CK does, which facilitates comparison with ATT&CK but may not intuitively describe the behavior of frontier AI systems or the main threats that they face. These dissimilarities arise for several reasons:

- **ATT&CK and ATLAS focus primarily on helping organizations determine how malicious actors could compromise their software systems, whereas an additional challenge for frontier model developers is determining how their models could impact the world directly** (e.g., if misused or given excessive autonomy). The closest match in ATLAS is the "System Misuse for External Effect" technique which is nested under the "Impact" tactic.[79] However, because this is relatively high-level, it may not be sufficient for frontier AI safety and security. For instance, frontier model developers might be interested in anticipating how their models could be used to create disinformation campaigns, conduct offensive cyber campaigns, or develop novel bioweapons.[80] Such information could help developers anticipate downstream stakeholders that they may need to engage, and to collaborate with them to develop appropriate countermeasures.

---

[78] "ATT&CK was created out of a need to systematically categorize adversary behavior as part of conducting structured adversary emulation exercises… The primary metric for success [of exercises] was 'How well are we doing at detecting documented adversary behavior?'" (Strom et al., 2018, p. 1).

[79] "Impact consists of techniques that adversaries use to disrupt availability or compromise integrity by manipulating business and operational processes" (MITRE ATLAS™, n.d.). As of August 2023, the techniques under "Impact" primarily focus on effects on the ML system itself, e.g., "Evade ML Model," "Erode ML Model Integrity," and "Denial of ML Service," although there is one technique, "System Misuse for External Effect," that covers potential misuse more generally.

- **Complex offensive cyber operations often move through distinct phases to establish increasing unauthorized access, while current attacks on ML systems can require relatively little build-up.** A cyber espionage campaign might involve an extended attempt to gain access to the network and establish a presence before finally taking action, reflected in several ATT&CK categories like "Initial Access," "Persistence," and "Defense Evasion." By contrast, many attacks on current ML systems do not require extended access to a model, and can be conducted via access to the public Application Programming Interface (API). For example, one of the most common attacks on large language models (LLMs) is prompt injection, which involves using an adversarially crafted input to cause a model to produce unexpected output without requiring unauthorized access to an organization's systems.[81] (However, it is unclear if this distinction will hold in the longer term, as frontier AI is a developing field and future attacks may be substantially more complex than currently observed.)

Though not strictly a threat-based approach, existing work on frontier model vulnerabilities could also inform a threat-based approach. The Open Worldwide Application Security Project (OWASP), an online community of security experts, has produced a list of the top ten most common LLM vulnerabilities that attackers can exploit in downstream applications (the "OWASP Top Ten for LLMs") (Wilson, 2023). Efforts to identify vulnerabilities in LLMs and other frontier models could make it easier to identify attack techniques that leverage these vulnerabilities, and to identify mitigations where these vulnerabilities cannot be robustly patched.[82] Existing work on frontier model capabilities will also likely inform a threat-based approach, as discussed in the "effect on world" section below.

## 5.3.2 | Proposed threat-based approaches

Frontier model developers should coordinate to develop a common taxonomy of tactics, techniques, and procedures (TTPs) covering two broad types of attacks:

1. **Effect on Model:** TTPs that malicious actors could use to manipulate models or elicit unwanted model behavior, e.g., by bypassing model guardrails.
2. **Effect on World:** TTPs that a malicious actor[83] could use to impact other actors and systems via successfully executing an "effect on model" attack or by using a model that has insufficient safeguards.

---

[80] E.g., see the concerns raised in the White House voluntary commitments (The White House, 2023b).
[81] In Wilson (2023), see "LLM01: Prompt Injections" on pp. 5-7.
[82] For example, researchers have demonstrated a general class of adversarial attacks on LLMs that can be developed via automated testing on open-source LLMs (e.g., Meta's LLaMa) and then used to target closed-source LLMs (e.g., ChatGPT, Bard, and Claude) (Zou et al., 2023). The transferability of these attacks makes them particularly difficult to guard against.
[83] Potentially, this could also include agent-based systems acting autonomously, as some experts have discussed (Hendrycks et al., 2023), but we treat this as a subset of malicious actors.

Below, we provide some illustrative examples of these threat-based frameworks, but these are not intended to be comprehensive as fully surveying all significant TTPs and developing a consensus taxonomy will require a sustained technical effort. If not done by MITRE, this effort could be taken up by a body such as the Frontier Model Forum, or through other channels between leading AI companies as agreed on in the July 2023 White House voluntary commitments.[84]

Knowledge base owners should strongly consider the need to limit public access to avoid facilitating attacks by malicious actors. They can consider sharing such information in settings with limited circulation, such as in the format of the Frontier Model Forum.

### An "effect on model" approach

The "effect on model" taxonomy could build on the existing ATLAS framework. However, to address the properties and vulnerabilities of machine learning models, we suggest that framework developers consider restructuring the 14 high-level tactics to better match the language and ontology used by machine learning practitioners rather than the ATT&CK tactics, which are targeted at security researchers.[85] Developing such a schema could help frontier model developers identify, prioritize, and mitigate possible attacks on their models.

To ensure the added material is relevant and comprehensive, framework developers should tap on existing resources documenting ML vulnerabilities, such as the OWASP Top Ten for LLMs and the OWASP Top Ten for ML Security, as well as reports by leading AI labs (OpenAI, 2023b; OWASP Foundation, 2023; Wilson, 2023). As an example, additional high-level tactics could include:

- **Compromise Training Pipeline:** The adversary is trying to manipulate model behavior by altering data or software that are used in the training of the model.[86] This could include altering third-party datasets that the developer would use in the pre-training phase (Dhar, 2023), providing malicious input during deployment to systems that use online learning,[87] or interfering with other software used during the training phase, including AI systems used to train the frontier model.[88]

- **Bypass Guardrails:** The adversary is trying to induce unauthorized model behavior that the developer has implemented safeguards to prevent. This could include various prompt injection techniques, which involve using adversarially crafted input to induce unauthorized outputs, such as providing harmful or offensive information.[89]

---

[84] In the detailed text of the commitments, companies agreed to "facilitate the sharing of information on advances in frontier capabilities and emerging risks and threats, such as attempts to circumvent safeguards" (The White House, 2023a). Developing a common taxonomy for a threat-based approach could facilitate and/or constitute such information sharing efforts.

[85] Reframing the ontology away from security researchers may conflict with ATLAS's goals, as the ATLAS website explicitly states: "We developed ATLAS to raise awareness of these threats and present them in a way familiar to security researchers" (*MITRE | ATLAS™, n.d.*). However, these two goals need not be mutually exclusive–framework developers could still map the revised ontology to the traditional ATT&CK framework to expedite use by security researchers.

[86] This is currently covered to some degree by, and might replace, "Persistence: The adversary is trying to maintain their foothold via machine learning artifacts or software," which includes the techniques "Backdoor ML Model" and "Poison Training Data." However, the two are arguably different; e.g., the Tay attack arguably did not require that attackers establish persistence on Microsoft's systems.

[87] Similar to how malicious users were able to attack Microsoft Tay.

[88] For example, OpenAI currently uses data classifiers to screen out some harmful data during the pre-training phase.

To make it easier for ML practitioners to use an effect-on-model framework, we also suggest that framework developers merge, re-order, and/or rename some of the current ATLAS categories; we footnote some such example suggestions, although a full review of such an effort is beyond the scope of this report.[90]

## An "effect on world" approach

Given present concern around misuse of frontier models, we believe it is also important to develop a common resource documenting TTPs for the malicious use of frontier models to attack other actors and systems. It could help labs and policymakers describe potential threats from frontier models using a common vocabulary, harden society against these threats, standardize risk assessments that feed into development and deployment decisions, and develop necessary countermeasures. Rather than just facilitating a defense-in-depth approach for individual organizations, such an approach might facilitate a defense-in-depth approach for society at large.

The high-level tactics for an "effect on world" taxonomy could draw from existing research to identify and evaluate potentially dangerous capabilities of frontier models, such as facilitating offensive cybersecurity campaigns, information operations, or weaponizable scientific research and manufacturing (Fist et al., 2023, Sec. 3.1; Shevlane et al., 2023, p. 5). The tactics should incorporate not just attacks that can be conducted with baseline frontier models, but also attacks that employ tools built with frontier models. For example, developers have built semi-autonomous agents using LLMs that can perform complex tasks without close supervision, which could lower the barrier for malicious actors wanting to do harm (Tong and Dastin, 2023).

Knowledge base owners should ensure that they decompose these tactics into more detailed techniques and procedures, grounding these in evidence from real-world observations and demonstrations, and what research suggests is possible.[91] For example, a breakdown of a "conduct offensive cybersecurity campaigns" tactic could include techniques such as conducting spear-phishing campaigns at scale (Hazell,

---

---

2023), evading detection through self-modification ([Sims, 2023](#)), or using autonomous agents to identify and exploit vulnerabilities,[92] all of which are techniques that have been demonstrated or seem feasible with additional research.

However, knowledge base owners should consider that there may be significant downside risks associated with publishing this information in a way that is broadly accessible to all members of the public. In frontier AI safety and security, the offense-defense balance could substantially favor the attacker–e.g., if vulnerabilities in frontier AI models are easy to exploit but difficult to patch, or attacks using frontier AI models (e.g., bioterror attacks) are easy to execute and difficult to develop countermeasures against ([Shevlane & Dafoe, 2019](#)). If so, it may be better for knowledge base owners to share information selectively, such as through the Frontier Model Forum. Existing practices in cybersecurity around coordinated vulnerability disclosure, where information about software vulnerabilities is shared selectively with product developers and vendors, and published publicly only on a time delay, could provide inspiration for responsible knowledge base development practices in frontier AI research.

### 5.3.3 | Application to national critical functions

From a policy perspective, governments could consider coupling this threat-based approach with an analysis of national critical functions (NCFs) that are vulnerable to attacks *on* AI systems, and attacks *by* AI systems. NCFs are "functions of government and the private sector so vital to the United States that their disruption, corruption, or dysfunction would have a debilitating effect on security, national economic security, national public health or safety, or any combination thereof," and the NCF construct is used by the US Cybersecurity and Infrastructure Agency (CISA) to "identify, analyze, prioritize, and manage" significant risks to these NCFs ([*National Critical Functions*, n.d.](#)).

Building on work done by industry to establish threats to and threats from frontier models, CISA or another research institution, such as RAND or MITRE, should develop frameworks to identify:

1. which NCFs will be most vulnerable to failures of robustness, resilience, safety, and security of ML models that they are using (i.e., "effect on model"), based on the projected extent of frontier model adoption by organizations supporting these NCFs; and
2. which NCFs will be most vulnerable to attacks using frontier models (i.e., "effect on world"), based on the projected capabilities of these models.

To an extent, such an effort could borrow methodologically from previous work done by RAND to assess risk to the NCFs from climate change, which was required by Executive Order (EO) 14008 ([Miro et al., 2022](#)). The RAND study assessed how 27 NCFs could be affected by climate change by 2030, by 2050, and by 2100 under two different greenhouse gas emissions scenarios. It identified (a) the NCFs at greatest risk, (b) the largest drivers of disruption, and (c) potential for cascading risk, and also conducted full risk assessments for the most vulnerable NCFs.[93]

---

[92] [Lohn et al. (2023)](#), pp. 24-25, also briefly discusses the use of autonomous offensive agents as a possible precursor to developing autonomous defensive agents.
[93] The study identified (a) the NCFs at greatest risk as Provide Public Safety and Supply Water, (b) the largest drivers of disruption as flooding, sea-level rise, and tropical cyclones and hurricanes, and (c) the Distribute Electricity NCF as having the highest potential for cascading risk in dependent NCFs ([Miro et al., 2022, p. vi](#)).

However, analyzing the impact of frontier AI on the NCFs will require a more dynamic approach than for climate change, given frontier AI's fast pace of progress. We recommend that the first version of such analysis be scoped tightly to a small subset (i.e., less than 10) of the most vulnerable NCFs, based on existing concerns around disinformation, cybersecurity, biosecurity, and the financial system.[94] Such analysis should emphasize detailed risk assessments for each of these NCFs, potential mitigations and technical countermeasures, and analysis of the residual risk after these mitigations are implemented.

We suggest that the analysis focus on a shorter time horizon (e.g., 5 to 10 years after initial publication), and treat this analysis as a "moving target" that will be dynamically reviewed and updated (e.g., every 1-2 years). The rapid pace of progress in frontier AI makes it infeasible to analyze frontier AI impacts on national critical functions over extended time periods (e.g., 30-70 years as in climate change), and static reports will become outdated. While the uncertainty around frontier AI progress will make developing projections difficult even over the 5-10 year timescale, it is also what could make such an effort especially valuable to provide strategic clarity on how frontier AI will affect national security.

## 5.4 | Limitations and future work

The threat-based approach faces several challenges and limitations:

- **Existing knowledge bases are not comprehensive for frontier AI systems and require expansion or restructuring.** The MITRE ATLAS database, OWASP Top Ten for LLMs, OWASP Top Ten for ML Security, and reports by leading AI labs present starting points for developing such an approach. However, currently, none of these are comprehensive for either "effect on model" or "effect on world" threats.

- **This approach focuses on adversarial actors and is less effective at anticipating or preventing several other classes of incidents.** For example, it is not effective at identifying robustness or resilience issues in critical infrastructure, or in identifying systemic social, economic, and political challenges precipitated by frontier AI systems.

- **Even after the threat databases are developed, it may be difficult for organizations to plan their defenses based on attacker TTPs if there is no mapping between attack techniques and defenses.** In the 2020 study of MITRE ATT&CK, only 57% of respondents said that ATT&CK was useful for helping to determine gaps in their security tools (Basra & Kaushik, 2020, p. 2).[95] Other results from the survey suggest that a key difficulty could be that these organizations are unable to map ATT&CK techniques to the security products they use, or the security events they observe.[96]

---

[94] For instance, the RAND study on climate change focuses only on 27 of the more vulnerable NCFs rather than all 55. We have discussed threats from disinformation, cybersecurity, and biosecurity elsewhere in this report; for risks to the financial system, see: Gensler & Bailey (2020); Sorkin et al., (2023).

[95] However, this implies that 43% of respondents did *not* see ATT&CK as useful for this purpose.

[96] From Basra & Kaushik (2020), p. 14: "Forty-five percent of organizations identified the lack of interoperability with security products while using ATT&CK, 43% cited the difficulty of mapping of event data to tactics and techniques, and 36% say they receive too many false positives."

To address these challenges, frontier AI developers should consider developing additional resources mapping attack techniques to potential defenses, such as MITRE Engenuity does for MITRE ATT&CK to NIST SP 800-53, or MITRE D3FEND does for a MITRE-developed set of security controls.[97]

- **The number of attack TTPs included can make high-level overviews and prioritization between TTPs difficult for users.** ATT&CK and ATLAS deliberately avoid assigning priorities to TTPs as such priorities are highly dependent on the context and risk profile of the organization.[98] This means that, unlike the functional or lifecycle approach, a threat-based approach as described above may not be suitable for senior stakeholders without significant work by the organization to prioritize and categorize the relevant TTPs and responses.

- **Aggregating and publicly distributing information about attacks on, or using, frontier AI models could enable malicious attacks if not done carefully.** Knowledge base owners should consider restricting the circulation of such information to actors responsible for addressing and mitigating these issues, and/or to delay public disclosure until these issues can be patched and mitigated.

# 6 | Evaluating and applying the suggested frameworks

We first evaluate and provide context for use of these frameworks, then explore how they can be applied to current measures proposed by labs.

## 6.1 | Context for applying frameworks

**We recommend that policymakers and frontier model developers adopt the functional framework first**, primarily given that this framework has the most existing infrastructure available (through NIST and CLTC). The functional framework is also relevant to stakeholders at all levels of seniority, given that it provides different levels of granularity ranging from the function, to category, to subcategory level. **Sophisticated users can also use the lifecycle framework to conduct gap analysis complementing the functional approach**, but because there is no comprehensive resource detailing the specific safety and security activities that should be conducted at each stage of the lifecycle, organizations will have to develop their own list of activities if they do so.

---

[97] E.g., see: _MITRE D3FEND Knowledge Graph_ (n.d.); _NIST 800-53 Control Mappings_, (n.d.).
[98] From Basra & Kaushik (2020), p. 15: "One other implementation issue we discovered from our survey results is that many organizations do not use the ATT&CK framework because it does not prioritize any adversary techniques, and no weights are assigned. We hypothesize that this is an intentional design decision, made to enable each business or security product to conduct its own independent risk assessment and identify which threats are more likely and have the greatest impact. Prioritization of tactics and techniques will be necessary for each enterprise based on the threat intelligence that they have for their sector and specific threat models."

**In the longer term, policymakers and frontier model developers should collaborate to develop detailed resources for all three frameworks.** Organizations should be aware that each framework has its strengths and weaknesses, and treat all three frameworks as overlapping and complementary.

Below, we provide a summary table capturing and comparing the main characteristics of these approaches.

| | Functional | Lifecycle | Threat-based |
|---|---|---|---|
| **Purpose** | Supports high-level risk management activities and resource allocation; assigns controls to lower-level outcomes. | Provides a holistic view of safety/security activities in software development, deployment, and operations. | Aids understanding of the motivations and methods of malicious actors in order to prepare effective defenses. |
| **Value proposition** | Ensures cross-cutting protections that provide resilience against known and unknown threats, even as technologies change. | Promotes "shift left" and "security by design"; calls attention to important deployment decisions and need for continuous monitoring. | Addresses adversarial actors; can facilitate whole-of-society defense against malicious use of frontier AI systems. |
| **Limitations** | Can be difficult to prioritize activities and evaluate coverage of threats. | Can omit cross-cutting categories of activities; does not focus on other actors' activities. | Exclusively focused on adversarial actors. |
| **Existing infrastructure that users can adopt for frontier AI risk management** | OK. Includes NIST AI RMF, CLTC risk profile for GPAI systems, and NIST CSF. | Limited. Various models exist, but no consensus and limited detail on specific activities. | Limited. MITRE ATLAS targets cyber experts, and has limited detail on frontier AI specific risks. |
| **Most suitable parties to conduct further research** | NIST, Frontier Model Forum | NIST, Frontier Model Forum | Frontier Model Forum, MITRE, CISA |

Ultimately, these approaches must collectively address a variety of risks and threats including:

- Robustness and/or resilience issues (e.g., failures of critical infrastructure or weapons systems);
- Misuse scenarios (e.g., creation of bioweapons);
- Novel classes of threats (e.g., threats from agentic systems);
- Systemic social and economic effects (e.g., degradation of media environment).

Individual approaches may be strong or weak against particular risks and threats–for example, threat-based approaches may be exceptionally good at anticipating and preventing misuse scenarios, but poor at most other tasks. Frontier AI developers should ensure that the combination of approaches is implemented in such a way that all of these categories of risks and threats are covered.

## 6.2 | Application to existing measures

At a high level, these frameworks can be used as a gap analysis tool to identify where companies need to bolster their defense-in-depth strategies. To illustrate, we apply the frameworks to a series of voluntary commitments, announced by the White House in July 2023, that seven leading AI companies agreed to (Amazon, Anthropic, Google, Inflection, Meta, Microsoft, and OpenAI) ([The White House, 2023b](#)). While these commitments were certainly not intended to be comprehensive, running through them is a useful exercise to demonstrate how the suggested defense-in-depth frameworks can help frontier AI developers and policymakers, and to test where the limitations of these frameworks are.[99] We limit our exercise to the functional and lifecycle frameworks, given the limited resources available for the threat-based approach.

The main commitments[100] that the seven companies agreed to are:

1. **Internal and external security testing** of AI systems before release, to guard against risks including biosecurity, cybersecurity, and broader societal effects.
2. **Information sharing on safety practices and attempts to subvert safeguards**, across industry and with other parties.
3. **Cybersecurity and insider threat safeguards** to protect model weights from being stolen.
4. **Vulnerability discovery and reporting mechanisms** that third parties can use after model release.
5. **Technical mechanisms to identify AI-generated content**, e.g., watermarking for audio and visual content.
6. **Public reporting of key model details**, such as capabilities, limitations, and areas of appropriate and inappropriate use.
7. **Research on societal risks**, including bias, discrimination, and privacy.

Based on this gap analysis exercise, we suggest some directions for future voluntary commitments:

- Establish and commit to governance practices that facilitate a culture of risk management, as outlined in NIST AI RMF Govern 1, Govern 2, and Govern 3.
- Commit to pre-deployment review mechanisms and establish best practices for pre-deployment review, as outlined in NIST AI RMF Manage 1 and Manage 2, and in the "Staged Deployment" phase of our proposed lifecycle model.

---

[99] These frameworks could also be used to classify proposed measures and identify areas for future work; for example, the Center for the Governance of AI (GovAI) conducted a survey asking experts to rate the usefulness of 50 proposed measures that AI labs could undertake. Further work could identify gaps in the 50 measures by benchmarking them against the above frameworks, or to group and prioritize measures within their respective categories ([Schuett et al., 2023](#)).

[100] We exclude one commitment, in which "the companies commit to develop and deploy advanced AI systems to help address society's greatest challenges," as it is relatively general and does not clearly address risks from AI systems.

- Establish and commit to best practices for post-deployment monitoring and incident response, as outlined in NIST AI RMF Manage 4, and establish standards for an effective monitoring and response regimen.[101]

A full analysis would also assess whether this approach meets the criteria that "no single layer, no matter how robust, is exclusively relied upon." Within each subcategory of the framework or activity of the lifecycle, frontier AI developers should ensure that there are multiple independent mechanisms that are resilient to each other's failure. As mentioned previously, the nascent state of AI safety and security and the brevity of the commitments means that we do not hold frontier AI developers to that standard in this gap analysis exercise.

## 6.2.1 | Functional

Below, we assign the commitments to the subcategories outlined in the NIST AI Risk Management Framework, to the extent possible given the limited information provided in the White House announcement and the detailed description of the commitments ([The White House, 2023a](#)). For brevity, we list the indicative subcategory by name and do not list the actual subcategory outcomes in full; desired subcategory outcomes are available in [Appendix B](#) for cross-reference if necessary.

1. Internal and external security testing **[Govern 2.1; Map 2.3, 4.1; Measure 1.1, 1.3, 2.3, 2.7]**[102]
2. Information sharing on safety practices and attempts to subvert safeguards **[Govern 4.2, 4.3, 5.1, 5.2]**
3. Cybersecurity and insider threat safeguards **[not captured in NIST AI RMF; better reflected by NIST Cybersecurity Framework]**
4. Vulnerability discovery and reporting mechanisms **[Measure 3.1, 3.3]**
5. Technical mechanisms to identify AI-generated content **[Govern 4.2; Measure 2.9; Manage 3.2]**[103]
6. Public reporting of key model details **[Govern 4.2, 4.3; Map 2.1, 2.2, 3.1, 3.2]**
7. Research on societal risks **[Govern 4.2; Map 4.1, 5.1, 5.2]**

The Map and Measure functions are relatively well-covered, with the commitments touching on all of their categories except for the "Measure 4" and "Map 1" categories.[104] This reflects the emphasis of the voluntary commitments on ways to characterize and stress-test frontier AI models and understand their impacts on national security and society. Some categories within the "Govern" function, particularly Govern 4 and Govern 5, are also fairly well-covered, reflecting the emphasis on information sharing with peer companies and the public.

---

[101] See "Deployment corrections: An incident response framework for frontier AI models" for details. ([O'Brien et al., 2023](#))

[102] These are aligned with a concept note by NIST to the NIST Generative AI Public Working Group, which links these specific subcategories to pre-deployment testing (red teaming). Description: "Pre-release and pre-deployment testing techniques can enable developers to map, measure and manage potential negative impacts prior to affecting users and consumers of AI technology." (National Institute of Standards and Technology, personal communication, August 8, 2023)

[103] These are aligned with a concept note by NIST to the NIST Generative AI Public Working Group, which links these specific subcategories to content provenance. Description: "Provenance techniques can enable users to identify if the content they are consuming is AI-generated or not. Provenance techniques also can enable individuals and organizations to trace protected content." (National Institute of Standards and Technology, personal communication, August 8, 2023)

[104] Specifically, as a crude way of assessing coverage, there are representative subcategories (e.g., "Measure 1.1") from each of the categories (e.g., "Measure 1") from Map 1 to Map 5, and from Measure 1 to Measure 3, although not Measure 4. Given the brevity of the commitments, it is to be expected that not all of the subcategories within each category will be covered.

However, the "Manage" function and several categories under "Govern" are noticeably absent, suggesting that future voluntary commitments–or regulation–could strengthen requirements in these areas. For example:

- **Frontier AI developers could commit to corporate governance practices that facilitate a culture of risk management** by enabling the other functions of mapping, measuring, and managing risk, as outlined in Govern 1, Govern 2, and Govern 3.[105] These subcategories describe measures that support an overall culture of risk management, such as developing transparent and effective processes, empowering staff to perform risk management duties, and ensuring accountability from leadership.

- **Frontier AI developers could commit to pre-deployment safeguards to manage risks during training and deployment decisions**, as described in Manage 1 and Manage 2. For example, researchers at UC Berkeley's Center for Long-Term Cybersecurity (CLTC) have recommended that in relation to Manage 1.1, frontier AI developers incorporate risk assessment results when making go/no-go decisions (Barrett, Newman, et al., 2023, pp 9-10). While the voluntary commitments discuss red teaming and testing extensively, they do not explicitly tie testing results to any deployment decisions.

- **Frontier AI developers could commit to post-deployment monitoring and incident reporting mechanisms**, as described in Manage 4. We describe such measures in a separate piece on "deployment corrections" (O'Brien et al., 2023).

The above suggestions are illustrative, and granular comparisons of the voluntary commitments with individual subcategories and associated supplementary guidance will provide further suggestions. For example, despite the strong emphasis on mapping and measuring risk in the voluntary commitments, frontier AI developers could still take further steps in the "Map" function to "set risk tolerance thresholds to prevent unacceptable risks," as the CLTC research group suggests as supplementary guidance under Map 1.5 (Barrett, Newman, et al., 2023, pp 9-10).

## 6.2.2 | Lifecycle

Below, we assign the commitments to the six broad phases identified in our proposed lifecycle framework. While lifecycle stage assignments are clearer for some commitments–e.g., "internal and external security testing"–some commitments cut across most or all of the AI model lifecycle.

1. Internal and external security testing **[Evaluate, Iterate, and Mitigate]**
2. Information sharing on safety practices and attempts to subvert safeguards **[multiple–cuts across lifecycle]**

---

[105] E.g., by adopting the three lines of defense structure outlined by Schuett (2022), or risk assessment processes outlined in Koessler & Schuett (2023).

3. Cybersecurity and insider threat safeguards **[multiple–cuts across lifecycle]**
4. Vulnerability discovery and reporting mechanisms **[Operate and Monitor]**
5. Technical mechanisms to identify AI-generated content **[Evaluate, Iterate, and Mitigate]**
6. Public reporting of key model details **[Staged Deployment]**
7. Research on societal risks **[Plan Scope and Design Architecture]**

These measures focus primarily on the latter half of the model lifecycle, particularly the category "Evaluate, Iterate, and Mitigate." These are important–e.g., the commitment to internal and external security testing could help to anticipate potentially dangerous behavior of frontier models.[106] Such testing is necessary to inform go/no-go decisions for model deployment.

However, there remains room for additional commitments at all phases of the model lifecycle. This is particularly true for safety and security activities earlier in the development cycle that aim to catch issues before late-stage testing and mitigation. This could include measures such as: software requirement specification techniques borrowed from safety-critical domains, dataset curation techniques, and foundational research to build safer and more secure AI systems. There is also room for further work on post-deployment monitoring and response, which we address in a separate publication (O'Brien et al., 2023).

# 7 | Conclusion

**Defense-in-depth as a principle is easy to understand, but detailed defense-in-depth strategies are difficult to get right.** In this report, we suggest three complementary frameworks–functional, lifecycle, and threat-based–that frontier AI developers and policymakers can use to ensure defenses against emerging risks from frontier AI are comprehensive and robust.

## 7.1 | Overview of Next Steps

We recommend that frontier AI developers and policymakers first adopt a functional approach using the NIST AI RMF, given the extensive infrastructure already developed or under development. But moving forward, developers and policymakers should work together to establish a detailed lifecycle model for frontier AI and build out a threat-based approach covering both an "effect on model" and "effect on world" approach. These approaches should be treated as complementary, as the complexity of the threat landscape associated with frontier AI systems means that no single framework can capture all possible issues.

**We have focused in this report on a defense-in-depth approach for *frontier AI developers*, but future work should also follow up on how to develop a defense-in-depth approach to mitigate AI risk for *society at large*.** In other words, rather than just describing how developers and policymakers can ensure that the model development and deployment process is conducted safely, a whole-of-society approach should consider what measures should be taken by semiconductor supply chain companies, downstream deployers such as critical infrastructure operators, homeland and national security authorities, other researchers and industry players, and possibly international agencies to guard against threats from frontier AI models.

---

[106] For example, see further discussion in Shevlane et al. (2023).

Such a strategy should involve not just preventing potential AI incidents, but also identifying ways to respond to and strengthen resilience against potential threats, including by developing tools that help build new defenses and reinforce existing institutions.[107] Developing such a strategy will require a sustained research effort, but could potentially be modeled after the Cyberspace Solarium Commission, which developed a "strategy of layered cyber deterrence" incorporating more than 80 recommendations, many of which have now become law (*Cyberspace Solarium Commission*, n.d.).[108]

## 7.2 | Recommendations

Below, we summarize our recommendations, along with the actors we believe are best suited to implement them. Identified actors include the Cybersecurity and Infrastructure Agency (CISA), the Department of Defense (DoD), the Department of Energy (DOE), the Department of Health and Human Services (HHS), the Frontier Model Forum (FMF), MITRE, the National Institute of Standards and Technology (NIST), the National Science Foundation (NSF), the Partnership on AI (PAI), the UC Berkeley Center for Long-Term Cybersecurity (CLTC), and other general categories of actors such as frontier AI developers, researchers, and philanthropists.

| Functional | |
|---|---|
| **1 \| Establish consensus on which categories of activities in the NIST AI RMF are the highest priority for frontier AI developers.** (3.3.1 \| The NIST AI RMF) NIST and/or the FMF, with researcher input, should identify high-priority categories for frontier AI safety and security. To ensure defense-in-depth, frontier AI developers should implement multiple independent measures for these categories. | NIST (or FMF), with researcher input (e.g., CLTC) |
| **2 \| Develop a detailed catalog of measures ("controls") that are important for frontier AI safety and security.** (3.3.3 \| Providing detailed controls) For instance, NIST SP 800-53 lists 1,000 detailed controls for cybersecurity across 20 "families." No current equivalent exists for AI, and it would be useful for frontier AI developers to have a similar catalog focused on frontier AI safety and security. | NIST, or industry bodies like FMF or PAI |

---

[107] For example, one expert involved in red-teaming GPT-4 has called for investment in "violet teaming": "identifying how a system (e.g., GPT-4) might harm an institution or public good, and then supporting the development of tools using that same system to defend the institution or public good" (Ovadya, 2023).

[108] See *Cyberspace Solarium Commission Executive Summary* (2020), p. 8 for description of the layered cyber deterrence strategy: "Shape Behavior, Deny Benefits, and Impose Costs." Lawmakers in the US House of Representative have introduced a bipartisan bill proposing a National AI Commission to draft a regulatory framework on AI, although it is unclear how this bill will interact with other competing efforts, including Senate Majority Leader Chuck Schumer's proposed regulatory framework (Sokler et al., 2023). While a National Security Commission on Artificial Intelligence (NSCAI) did exist and has since concluded its work (succeeded by the Special Competitive Studies Project, or SCSP), the NSCAI and SCSP have focused primarily on strengthening US competitiveness in AI and other technologies against other adversaries such as China, rather than on regulating threats from frontier models per se.

| Lifecycle | |
|---|---|
| **3 \| Establish a detailed lifecycle framework for frontier AI that describes safety and security activities at each stage.** (4.3.2 \| [Proposed lifecycle framework](#)) This framework can build on work by the OECD while incorporating details from frontier AI developers, and should map activities to the NIST AI RMF where possible. It should ensure all phases are appropriately covered, which could include a "shift left" (see recommendation 4), and a stage for post-deployment monitoring and response. | FMF and/or NIST |
| **4 \| Pursue research that supports a "shift left" for frontier AI by emphasizing safety and security activities earlier in the development cycle.** (4.3.3.1 \| ["Shifting left" on AI safety and security](#); 6.2.2 \| [Lifecycle](#)) Potential research areas could include: software requirement specification techniques borrowed from safety-critical domains, dataset curation techniques, and foundational research to build safer and more secure AI systems. | Frontier AI developers, philanthropists, and major government funders of AI R&D (e.g., the NSF, DoD, DOE, and HHS) |
| **Threat-based** | |
| **5 \| Restructure and expand MITRE ATLAS to further address attacks on frontier AI.** (5.3.2.1 \| [An "effect on model" approach](#)) MITRE ATLAS is a knowledge base of tactics, techniques, and procedures (TTPs) that malicious actors can use to attack AI systems. The high-level categories ("tactics") are closely adapted from the equivalent cybersecurity knowledge base. We suggest restructuring these high-level tactics to reflect an AI-specific taxonomy (e.g., to include tactics like compromising training pipelines), and expanding on techniques and procedures that could enable misuse such as bypassing model guardrails. | FMF, MITRE, and/or frontier AI developers |
| **6 \| Develop a common taxonomy of TTPs describing malicious use of frontier models to impact other actors and systems.** (5.3.2.2 \| [An "effect on world" approach](#)) The knowledge base should combine real-world evidence and what research suggests is possible. Database owners should strongly consider limiting public access, due to the risk of facilitating attacks by malicious actors. | FMF, MITRE, and/or frontier AI developers |
| **7 \| Establish a mechanism to assess and monitor potential effects of frontier AI systems on the top ten most vulnerable National Critical Functions.** (5.3.3 \| [Application to national critical functions](#)) These effects should be re-evaluated at least once every 1-2 years, and should be informed by the "effect on model" and "effect on world" databases described in recommendations 5 and 6. | CISA |

# Acknowledgments

We are grateful to the following people for providing valuable feedback and insights: Onni Aarne, Ashwin Acharya, Bill Anderson-Samways, Renan Araujo, Haydn Belfield, Asher Brass, Marie Buhl, Siméon Campos, Ben Cottier, Samuel Curtis, Bill Drexel, Tim Fist, Matt Gentzel, Jason Green-Lowe, Erich Grunewald, Oliver Guest, Maia Hamin, Vance Hilderman, Hamish Hobbs, Caroline Jeanmaire, Leonie Koessler, Noam Kolt, Jam Kraprayoon, Yolanda Lannquist, Jung-Ju Lee, Patrick Levermore, Morgan Livingston, Sebastian Lodemann, Jon Menaster, Nikhil Mulani, Malcolm Murray, Richard Mallah, Nicolas Moës, Abi Olvera, Devansh Pandey, Robert Praas, Max Räuker, Morgan Simpson, Ben Snodin, Zach Stein-Perlman, Helen Toner, Peter Wildeford, Emma Williamson, Caleb Withers, and Jenny Xiao. Special thanks to Anthony Barrett from UC Berkeley's Center for Long-Term Cybersecurity for taking the time to engage on the report and his closely related supplementary guidance for the NIST AI RMF; Luke Muehlhauser for providing a starting point for this piece by suggesting "defense-in-depth" be applied to AI and sharing his initial notes; Jonas Schuett for several invaluable critiques that helped us make this piece more timely and relevant; and Paul Scharre for generously offering to host a discussion of this report at the Center for a New American Security. We are also grateful to Adam Papineau for copy editing and Umar Rafique for graphic design. All errors are our own.

# Appendix A: Relevant frameworks in nuclear reactor safety and cybersecurity

We reviewed several relevant concepts and frameworks from other domains–principally nuclear reactor safety and cybersecurity–that we were not able to detail in full due to time constraints and their lesser relevance. Here we provide a non-comprehensive overview of these to facilitate future research.

## Appendix A-1: Defense-in-depth levels in nuclear reactor safety

Defense-in-depth in nuclear reactor safety is often formalized as a series of "levels" (or "layers"), ranging from two to five in number, corresponding to different stages in time and severity as a nuclear incident or accident develops.[109] The table below is taken from a relatively authoritative treatment of defense-in-depth, namely a 1996 publication by the International Nuclear Safety Advisory Group (INSAG), which is convened by the International Atomic Energy Agency (IAEA).

| Levels of defense in depth | Objective | Essential means |
|---|---|---|
| Level 1 | Prevention of abnormal operation and failures | Conservative design and high quality in construction and operation |
| Level 2 | Control of abnormal operation and detection of failures | Control, limiting and protection systems, and other surveillance features |
| Level 3 | Control of accidents within the design basis | Engineered safety features and accident procedures |
| Level 4 | Control of severe plant conditions, including prevention of accident progression and mitigation of the consequences of severe accidents | Complementary measures and accident management |
| Level 5 | Mitigation of radiological consequences of significant releases of radioactive materials | Off-site emergency response |

*Source: International Nuclear Safety Advisory Group (1996), Table 1 (p. 6) on "Levels of Defense in Depth."*

## Appendix A-2: Relevant cybersecurity frameworks

### Defense-in-depth frameworks

---

[109] See Drouin et al. (2016), p. 207: "there is no agreement in the number of layers of defense. They vary from two layers, prevention and mitigation, to five layers" depending on the national or international agency consulted.

This overview is meant to be illustrative of the variety of definitions of defense-in-depth, rather than exhaustive.

- **People, technology, and operations (or processes):** Several NIST publications define defense-in-depth as "an information security strategy that integrates people, technology, and operations capabilities to establish variable barriers across multiple layers and missions of the organization."[110] The National Security Agency (NSA) and Department of Defense (DoD) have used this definition since at least the early 2000s.[111] Some other publications refer to "processes" rather than "operations."[112]

- **Layering defenses by network zone:** Some organizational networks can be separated into multiple network zones of increasing importance. For example, a Department of Homeland Security guide for protecting industrial control systems (e.g., to control oil/gas pipelines) describes four network zones of increasing sensitivity: Zone 1 providing external connectivity to the Internet, Zone 2 for corporate communications, and Zone 3 and 4 for control systems and their communications.[113] Not all networks can be subdivided this way, especially contemporary ones.

- **Layering different types of security measures ("controls"):** A number of defense-in-depth descriptions simply involve dividing controls into multiple categories; "depth" derives from the inclusion of controls from multiple such categories. Some contemporary industry sources, including the official guide to a popular cybersecurity certification, divide a defense-in-depth approach into "physical, technical, and administrative" controls.[114] Another cybersecurity playbook by Carnegie Mellon University's Software Engineering Institute from 2006 identifies eight categories of controls: compliance, risk, identity, authorization, accountability, availability,

---

[110] See CSRC (n.d.); this is also the definition used by NIST SP 800-53 Rev. 5 (Joint Task Force, 2020). Note that while this conveys the need for layered defense, it does not emphasize the need to avoid single points of failure as strongly as nuclear reactor safety does.

[111] A 2004 document by the NSA states: "An important principle of the Defense in Depth strategy is that achieving Information Assurance requires a balanced focus on three primary elements: People, Technology and Operations" (US National Security Agency, 2004). DoD Directive 8500.01, issued in October 2002, also defines defense-in-depth as "The DoD approach for establishing an adequate IA posture in a shared-risk environment that allows for shared mitigation through: the integration of people, technology, and operations; the layering of IA solutions within and among IT assets; and, the selection of IA solutions based on their relative level of robustness." (US Department of Defense, 2002, p. 18).

[112] "Defence in depth is the intelligent security management of people, processes and technology, in a holistic risk-management approach" (*Defence in Depth*, 2008, p. 6).

[113] See Idaho National Laboratory Control Systems Security Center (2006), pp. 15-17: "Isolating and Protecting Assets: Defense-in-Depth Strategies." While this document does not explicitly define defense-in-depth, it frequently refers to using network zones to create depth, e.g., on p. 17: "Thus, defensive strategies that secure each of the core zones can create a defensive strategy with depth, offering the administrators more opportunities for information and resources control, as well as introducing cascading countermeasures that will not necessarily impede business functionality."

[114] Figure 14.1 of the CISSP (ISC)2 Certified Information Systems Security Professional Official Study Guide (Stewart et al., 2015) describes "defense in depth with layered security" referring to physical access controls, logical/technical controls, and administrative access controls. Physical controls refer to barriers to stop physical access, technical controls to network and system barriers that stop digital access, and administrative barriers to policies and governance processes. Other industry sources include Fruhlinger (2022): "One way of thinking about defense in depth as a whole groups defensive elements into three main categories: administrative controls, physical controls, and technical controls." and Chancey (2019): "Defense in Depth is simply defined as having security controls in more than one of the three areas of security. Generally, the three areas are regarded as Administrative Controls, Physical Controls, and Technical Controls."

configuration, and incident management.[115]

Defense-in-depth also now coexists in cybersecurity with other concepts, particularly "zero trust" and "assume breach." "Zero trust" network architecture emerged around 2010 around the assumption that "there is no implicit trust granted to assets or user accounts based solely on their physical or network location... or based on asset ownership" ([Rose et al., 2020](#)). It links closely to the "assume breach" mindset, which suggests that organizations should design defenses with the assumption that an attacker is already inside their systems.[116]

Some practitioners frame zero-trust and "assume breach" as opposed to a version of defense-in-depth that takes depth as network depth and emphasizes protecting the network perimeter. However, we use defense-in-depth more broadly, and so view these concepts as complementary.

## NIST SP 800-172: Defense-in-depth against advanced persistent threats

Another recent prominent example of defense-in-depth in cybersecurity is the strategy laid out in the National Institute of Standards and Technology Special Publication 800-172, or NIST SP 800-172 ([Information Technology Laboratory Computer Security Division, 2021](#)). NIST SP 800-172 is a compilation of security controls that organizations can adopt to protect sensitive information from advanced persistent threats (APTs), common parlance in cybersecurity for nation-state actors.[117] It frames these controls within a "defense-in-depth protection strategy" that has three main thrusts: (1) penetration-resistant architecture, (2) damage-limiting operations, and (3) designing for cyber resiliency and survivability.[118]

- **Penetration-resistant architecture** is used to "limit the opportunities for an adversary to compromise an organizational system and to achieve a persistent presence in the system." This includes elements like controlling information flows between security domains (3.1.3e), automating the inventory of system components and detection of misconfigured components (3.4.2e and 3.4.3e), software supply chain risk assessment and management (3.11.6e and 3.11.7e), etc.
- **Damage-limiting operations** are used to "maximize the ability of an organization to detect successful system compromises by an adversary and to limit the effects of such compromises".

---

[115] For a 2006 CMU report on defense-in-depth in cybersecurity, see: [May et al. (2006)](#).

[116] On "assume breach", see *[Embracing a Zero Trust Security Model](#)* (2021): "Consciously operate and defend resources with the assumption that an adversary already has presence within the environment. Deny by default and heavily scrutinize all users, devices, data flows, and requests for access. Log, inspect, and continuously monitor all configuration changes, resource accesses, and network traffic for suspicious activity." See also [Rose et al. (2020)](#), p. 8, which does not explicitly name the "assume breach" mindset but lists the first of six assumptions around ZTA as: "The entire enterprise private network is not considered an implicit trust zone. Assets should always act as if an attacker is present on the enterprise network."

[117] Specifically, NIST SP 800-172 specifically lays out counter-APT controls for nonfederal organizations that need to protect Controlled Unclassified Information (CUI). It complements NIST SP 800-171, which lays out generic controls that nonfederal organizations need to protect CUI (from threats that may not include APTs). Both are built on top of NIST SP 800-53, which provides an extended catalog of controls.

[118] "The enhanced security requirements provide the foundation for a multidimensional, defense-in-depth protection strategy through (1) penetration-resistant architecture, (2) damage-limiting operations, and (3) designing for cyber resiliency and survivability that support and reinforce one another" ([Information Technology Laboratory Computer Security Division, 2021](#)).

This includes elements like maintaining a Security Operations Center (3.6.1e) and cybersecurity incident response team (3.6.2e), threat hunting (3.11.2e), penetration testing (3.12.1e), etc.

- **Designing for cyber resiliency and survivability** is used to "prepare for, withstand, recover from, and adapt to compromises of cyber resources in order to maximize mission or business operations." This includes elements like using a diverse range of system components to limit malicious code propagation (3.13.1e), changing systems and system components to introduce unpredictability (3.13.2e), employing technical and procedural methods to confuse and mislead adversaries (3.13.3e), etc.

One of NIST SP 800-172's strengths is its recognition that a determined adversary will likely be able to breach an organization's perimeter defenses, and must therefore also take additional steps to "outmaneuver, confuse, deceive, mislead, and impede the adversary" once the adversary is in the defender's systems, so that the defenders can protect their "critical programs and high value assets."[119] This is reflected in the balance between the three main thrusts of the strategy, each of which broadly address different aspects of a breach.

## Appendix A-3: The NIST Cybersecurity Framework (CSF)

In cybersecurity, the NIST Cybersecurity Framework (CSF) covers five functions: Identify, Protect, Detect, Respond, and Recover (IPDRR). NIST CSF 1.0 was originally developed to protect US critical infrastructure, but has since been adapted by both governments and companies globally.[120] At the time of writing, it has been updated to version 1.1, but this version is currently under revision and will be re-released as NIST CSF 2.0 in early 2024. NIST CSF 2.0 will add a new "Govern" function but otherwise maintain the five-function IPDRR framework.[121] As NIST CSF 2.0 is not yet finalized, we do not describe the "Govern" function here.

NIST CSF 1.1 ([National Institute of Standards and Technology, 2018, pp. 7–8](#)) describes the five "core functions" as:

1. **Identify:** Develop an organizational understanding to manage cybersecurity risk to systems, people, assets, data, and capabilities.

---

[119] "This strategy recognizes that, despite the best protection measures implemented by organizations, the APT may find ways to breach primary boundary defenses and deploy malicious code within a defender's system. When this situation occurs, organizations must have access to additional safeguards and countermeasures to outmaneuver, confuse, deceive, mislead, and impede the adversary—that is, to take away the adversary's tactical advantage and protect and preserve the organization's critical programs and high value assets" ([Information Technology Laboratory Computer Security Division, 2021](#)).

[120] "While the CSF was originally developed to address the cybersecurity risks of critical infrastructure first and foremost, it has since been used much more widely" (*[NIST Cybersecurity Framework 2.0 Concept Paper: Potential Significant Updates to the Cybersecurity Framework, 2023, p. 4](#)*).

[121] "Reflecting substantial input to NIST, CSF 2.0 will include a new "Govern" Function to emphasize cybersecurity risk management governance outcomes… This new crosscutting Function will highlight that cybersecurity governance is critical to managing and reducing cybersecurity risk. Cybersecurity governance may include determination of priorities and risk tolerances of the organization, customers, and larger society; assessment of cybersecurity risks and impacts; establishment of cybersecurity policies and procedures; and understanding of cybersecurity roles and responsibilities" (*[NIST Cybersecurity Framework 2.0 Concept Paper: Potential Significant Updates to the Cybersecurity Framework, 2023, p. 10](#)*).

2. **Protect:** Develop and implement appropriate safeguards to ensure delivery of critical services.
3. **Detect:** Develop and implement appropriate activities to identify the occurrence of a cybersecurity event.
4. **Respond:** Develop and implement appropriate activities to take action regarding a detected cybersecurity incident.
5. **Recover:** Develop and implement appropriate activities to maintain plans for resilience and to restore any capabilities or services that were impaired due to a cybersecurity incident.

These functions "are not intended to form a serial path," and NIST suggests that organizations should perform them "concurrently and continuously." To operationalize the five functions, each is split up into multiple "outcome categories." For example, "Identify" is subdivided into: Asset Management; Business Environment; Governance; Risk Assessment; and Risk Management Strategy. The outcome categories can in turn be connected to even more specific activities, standards, guidelines, and practices, but organizations should exercise their own judgment in deciding which of these more granular measures to adopt (National Institute of Standards and Technology, 2018, p. 7).

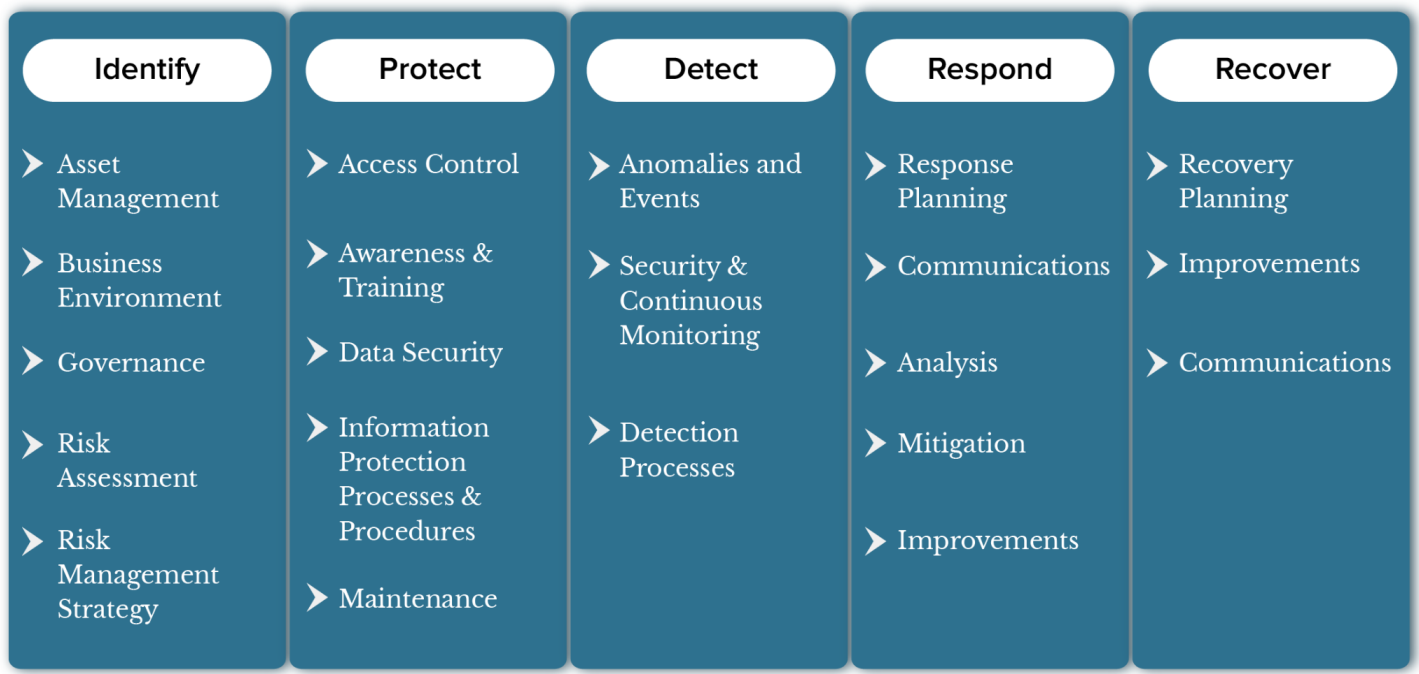

*Source: NIST Cybersecurity Framework, version 1.1. Table redesigned.*

## Common uses of the NIST CSF

Identifying top-level categories of activities can "aid organizations in easily expressing their management of... risk at a high level and enabling risk management decisions" ("The Five Functions," 2018). Senior decision-makers can use these functions as a "dashboard" to provide an overview of what measures they have in place for each of the main functions, and assess qualitatively whether these measures are meeting the desired outcomes for each function. This can facilitate decisions whether to bolster certain functions or to pare measures back in others.

Organizations can also calculate their financial spending on each function and use this to inform their budget allocation for risk management. For example, the US federal government's annual review of information technology and cybersecurity funding typically summarizes spending across all federal government agencies (excluding the US Department of Defense) in the five NIST framework functions.[122]

**NIST Framework Function Civilian CFO Act Agency Funding Totals (FY 2022, in millions of dollars)**

| NIST Function | Identify | Protect | Detect | Respond | Recover |
|---|---|---|---|---|---|
| FY 2022 spend (USD, millions) | $2,894 | $3,622 | $1,108 | $1,488 | $290 |
| % of total | 30.8% | 38.5% | 11.8% | 15.8% | 3.1% |

*Source: (Biden, 2021) (adapted); note that this excludes US Department of Defense spending.*

While this can provide an intuitive sense of the balance of costs between each function and the investment required to achieve a given set of outcomes, organizations should be careful about focusing on financial inputs without considering outcomes. While it may be easier to count the financial spending or number of measures allocated to a given function, this does not ultimately reflect whether the outcomes described by the function are being meaningfully achieved. Directly comparing inputs between functions can also be misleading if the organization does not take outcomes into account.[123]

# Appendix B: NIST AI Risk Management Framework

As we frequently reference the NIST AI RMF in this report (particularly in Section 3 and Section 6), below we reproduce a summary of the key categories and subcategories from the NIST AI RMF (Tabassi, 2023, pp. 20-32). While safety and security risks are among those that the NIST AI RMF addresses, it also covers other risks such as validity and reliability, accountability and transparency, explainability and interpretability, privacy, and fairness. It is also designed for relevance to many stakeholders including developers of smaller or sector-specific models and downstream users, not purely frontier AI models.

Additional guidance is available both in the NIST AI RMF, and in the NIST AI RMF Playbook (NIST AIRC Team, n.d.-b).

## Appendix B-1: Govern

**Under the "Govern" function, "a culture of risk management is cultivated and present."**

---

[122] E.g., see Table 12-3, "NIST Framework Function Civilian CFO Act Agency Funding Totals," in Biden (2021).
[123] For example, in the NIST CSF, the high cost of infrastructure defenses, which tend to be hardware-based, can drive up the costs of the "Protect" function–as seen in the FY2022 budget described above, where the US federal government allots 38.5% of its total spending to Protect, compared to 51.5% on all the four other functions (Biden, 2021).

Govern 1: Policies, processes, procedures, and practices across the organization related to the mapping, measuring, and managing of AI risks are in place, transparent, and implemented effectively.

- Govern 1.1: Legal and regulatory requirements involving AI are understood, managed, and documented.
- Govern 1.2: The characteristics of trustworthy AI are integrated into organizational policies, processes, and procedures.
- Govern 1.3: Processes and procedures are in place to determine the needed level of risk management activities based on the organization's risk tolerance.
- Govern 1.4: The risk management process and its outcomes are established through transparent policies, procedures, and other controls based on organizational risk priorities.
- Govern 1.5: Ongoing monitoring and periodic review of the risk management process and its outcomes are planned, organizational roles and responsibilities are clearly defined, including determining the frequency of periodic review.
- Govern 1.6: Mechanisms are in place to inventory AI systems and are resourced according to organizational risk priorities.
- Govern 1.7: Processes and procedures are in place for decommissioning and phasing out of AI systems safely and in a manner that does not increase risks or decrease the organization's trustworthiness.

Govern 2: Accountability structures are in place so that the appropriate teams and individuals are empowered, responsible, and trained for mapping, measuring, and managing AI risks.

- Govern 2.1: Roles and responsibilities and lines of communication related to mapping, measuring, and managing AI risks are documented and are clear to individuals and teams throughout the organization.
- Govern 2.2: The organization's personnel and partners receive AI risk management training to enable them to perform their duties and responsibilities consistent with related policies, procedures, and agreements.
- Govern 2.3: Executive leadership of the organization takes responsibility for decisions about risks associated with AI system development and deployment.

Govern 3: Workforce diversity, equity, inclusion, and accessibility processes are prioritized in the mapping, measuring, and managing of AI risks throughout the lifecycle.

- Govern 3.1: Decision-making related to mapping, measuring, and managing AI risks throughout the lifecycle is informed by a diverse team (e.g., diversity of demographics, disciplines, experience, expertise, and backgrounds).
- Govern 3.2: Policies and procedures are in place to define and differentiate roles and responsibilities for human-AI configurations and oversight of AI systems.

Govern 4: Organizational teams are committed to a culture that considers and communicates AI risk.

- Govern 4.1: Organizational policies and practices are in place to foster a critical thinking and safety-first mindset in the design, development, deployment, and uses of AI systems to minimize negative impacts.
- Govern 4.2: Organizational teams document the risks and potential impacts of the AI technology they design, develop, deploy, evaluate and use, and communicate about the impacts more broadly.

- Govern 4.3: Organizational practices are in place to enable AI testing, identification of incidents, and information sharing.

Govern 5: Processes are in place for robust engagement with relevant AI actors.
- Govern 5.1: Organizational policies and practices are in place to collect, consider, prioritize, and integrate feedback from those external to the team that developed or deployed the AI system regarding the potential individual and societal impacts related to AI risks.
- Govern 5.2: Mechanisms are established to enable AI actors to regularly incorporate adjudicated feedback from relevant AI actors into system design and implementation.

Govern 6: Policies and procedures are in place to address AI risks and benefits arising from third-party software and data and other supply chain issues.
- Govern 6.1: Policies and procedures are in place that address AI risks associated with third-party entities, including risks of infringement of a third party's intellectual property or other rights.
- Govern 6.2: Contingency processes are in place to handle failures or incidents in third-party data or AI systems deemed to be high-risk.

## Appendix B-2: Map

**Under the "Map" function, "context is recognized and risks related to context are identified."**

Map 1: Context is established and understood.
- Map 1.1: Intended purpose, potentially beneficial uses, context-specific laws, norms and expectations, and prospective settings in which the AI system will be deployed are understood and documented. Considerations include: specific set or types of users along with their expectations; potential positive and negative impacts of system uses to individuals, communities, organizations, society, and the planet; assumptions and related limitations about AI system purposes; uses and risks across the development or product AI lifecycle; TEVV and system metrics.
- Map 1.2: Inter-disciplinary AI actors, competencies, skills and capacities for establishing context reflect demographic diversity and broad domain and user experience expertise, and their participation is documented. Opportunities for interdisciplinary collaboration are prioritized.
- Map 1.3: The organization's mission and relevant goals for the AI technology are understood and documented.
- Map 1.4: The business value or context of business use has been clearly defined or – in the case of assessing existing AI systems – re-evaluated.
- Map 1.5: Organizational risk tolerances are determined and documented.
- Map 1.6: System requirements (e.g., "the system shall respect the privacy of its users") are elicited from and understood by relevant AI actors. Design decisions take socio-technical implications into account to address AI risks.

Map 2: Categorization of the AI system is performed.
- Map 2.1: The specific task, and methods used to implement the task, that the AI system will support is defined (e.g., classifiers, generative models, recommenders).
- Map 2.2: Information about the AI system's knowledge limits and how system output may be utilized and overseen by humans is documented. Documentation provides sufficient information to assist relevant AI actors when making informed decisions and taking subsequent actions.

- Map 2.3: Scientific integrity and TEVV considerations are identified and documented, including those related to experimental design, data collection and selection (e.g., availability, representativeness, suitability), system trustworthiness, and construct validation.

Map 3: AI capabilities, targeted usage, goals, and expected benefits and costs compared with appropriate benchmarks are understood.
- Map 3.1: Potential benefits of intended AI system functionality and performance are examined and documented.
- Map 3.2: Potential costs, including non-monetary costs, which result from expected or realized AI errors or system functionality and trustworthiness - as connected to organizational risk tolerance - are examined and documented.
- Map 3.3: Targeted application scope is specified and documented based on the system's capability, established context, and AI system categorization.
- Map 3.4: Processes for operator and practitioner proficiency with AI system performance and trustworthiness – and relevant technical standards and certifications – are defined, assessed and documented.
- Map 3.5: Processes for human oversight are defined, assessed, and documented in accordance with organizational policies from GOVERN function.

Map 4: Risks and benefits are mapped for all components of the AI system including third-party software and data.
- Map 4.1: Approaches for mapping AI technology and legal risks of its components – including the use of third-party data or software – are in place, followed, and documented, as are risks of infringement of a third-party's intellectual property or other rights.
- Map 4.2: Internal risk controls for components of the AI system including third-party AI technologies are identified and documented.

Map 5: Impacts to individuals, groups, communities, organizations, and society are characterized.
- Map 5.1: Likelihood and magnitude of each identified impact (both potentially beneficial and harmful) based on expected use, past uses of AI systems in similar contexts, public incident reports, feedback from those external to the team that developed or deployed the AI system, or other data are identified and documented.
- Map 5.2: Practices and personnel for supporting regular engagement with relevant AI actors and integrating feedback about positive, negative, and unanticipated impacts are in place and documented.

## Appendix B-3: Measure

**Under the "Measure" function, "identified risks are assessed, analyzed, or tracked."**

Measure 1: Appropriate methods and metrics are identified and applied.
- Measure 1.1: Approaches and metrics for measurement of AI risks enumerated during the Map function are selected for implementation starting with the most significant AI risks. The risks or trustworthiness characteristics that will not – or cannot – be measured are properly documented.
- Measure 1.2: Appropriateness of AI metrics and effectiveness of existing controls is regularly assessed and updated including reports of errors and impacts on affected communities.

- Measure 1.3: Internal experts who did not serve as front-line developers for the system and/or independent assessors are involved in regular assessments and updates. Domain experts, users, AI actors external to the team that developed or deployed the AI system, and affected communities are consulted in support of assessments as necessary per organizational risk tolerance.

Measure 2: AI systems are evaluated for trustworthy characteristics.
- Measure 2.1: Test sets, metrics, and details about the tools used during test, evaluation, validation, and verification (TEVV) are documented.
- Measure 2.2: Evaluations involving human subjects meet applicable requirements (including human subject protection) and are representative of the relevant population.
- Measure 2.3: AI system performance or assurance criteria are measured qualitatively or quantitatively and demonstrated for conditions similar to deployment setting(s). Measures are documented.
- Measure 2.4: The functionality and behavior of the AI system and its components – as identified in the MAP function – are monitored when in production.
- Measure 2.5: The AI system to be deployed is demonstrated to be valid and reliable. Limitations of the generalizability beyond the conditions under which the technology was developed are documented.
- Measure 2.6: AI system is evaluated regularly for safety risks – as identified in the MAP function. The AI system to be deployed is demonstrated to be safe, its residual negative risk does not exceed the risk tolerance, and can fail safely, particularly if made to operate beyond its knowledge limits. Safety metrics implicate system reliability and robustness, real-time monitoring, and response times for AI system failures.
- Measure 2.7: AI system security and resilience – as identified in the MAP function – are evaluated and documented.
- Measure 2.8: Risks associated with transparency and accountability – as identified in the MAP function – are examined and documented.
- Measure 2.9: The AI model is explained, validated, and documented, and AI system output is interpreted within its context – as identified in the MAP function – and to inform responsible use and governance.
- Measure 2.10: Privacy risk of the AI system – as identified in the MAP function – is examined and documented.
- Measure 2.11: Fairness and bias – as identified in the MAP function – is evaluated and results are documented.
- Measure 2.12: Environmental impact and sustainability of AI model training and management activities – as identified in the MAP function – are assessed and documented.
- Measure 2.13: Effectiveness of the employed TEVV metrics and processes in the MEASURE function are evaluated and documented.

Measure 3: Mechanisms for tracking identified AI risks over time are in place.
- Measure 3.1: Approaches, personnel, and documentation are in place to regularly identify and track existing, unanticipated, and emergent AI risks based on factors such as intended and actual performance in deployed contexts.
- Measure 3.2: Risk tracking approaches are considered for settings where AI risks are difficult to assess using currently available measurement techniques or where metrics are not yet available.

- Measure 3.3: Feedback processes for end users and impacted communities to report problems and appeal system outcomes are established and integrated into AI system evaluation metrics.

Measure 4: Feedback about efficacy of measurement is gathered and assessed.
- Measure 4.1: Measurement approaches for identifying AI risks are connected to deployment context(s) and informed through consultation with domain experts and other end users. Approaches are documented.
- Measure 4.2: Measurement results regarding AI system trustworthiness in deployment context(s) and across AI lifecycle are informed by input from domain experts and other relevant AI actors to validate whether the system is performing consistently as intended. Results are documented.
- Measure 4.3: Measurable performance improvements or declines based on consultations with relevant AI actors including affected communities, and field data about context-relevant risks and trustworthiness characteristics, are identified and documented.

## Appendix B-4: Manage

**Under the "Manage" function, "risks are prioritized and acted upon based on a projected impact."**

Manage 1: AI risks based on assessments and other analytical output from the MAP and MEASURE functions are prioritized, responded to, and managed.
- Manage 1.1: A determination is as to whether the AI system achieves its intended purpose and stated objectives and whether its development or deployment should proceed.
- Manage 1.2: Treatment of documented AI risks is prioritized based on impact, likelihood, or available resources or methods.
- Manage 1.3: Responses to the AI risks deemed high priority as identified by the Map function, are developed, planned, and documented. Risk response options can include mitigating, transferring, avoiding, or accepting.
- Manage 1.4: Negative residual risks (defined as the sum of all unmitigated risks) to both downstream acquirers of AI systems and end users are documented.

Manage 2: Strategies to maximize AI benefits and minimize negative impacts are planned, prepared, implemented, documented, and informed by input from relevant AI actors.
- Manage 2.1: Resources required to manage AI risks are taken into account, along with viable non-AI alternative systems, approaches, or methods – to reduce the magnitude or likelihood of potential impacts.
- Manage 2.2: Mechanisms are in place and applied to sustain the value of deployed AI systems.
- Manage 2.3: Procedures are followed to respond to and recover from a previously unknown risk when it is identified.
- Manage 2.4: Mechanisms are in place and applied, responsibilities are assigned and understood to supersede, disengage, or deactivate AI systems that demonstrate performance or outcomes inconsistent with intended use.

Manage 3: AI risks and benefits from third-party entities are managed.
- Manage 3.1: AI risks and benefits from third-party resources are regularly monitored, and risk controls are applied and documented.

- Manage 3.2: Pre-trained models which are used for development are monitored as part of AI system regular monitoring and maintenance.

Manage 4: Risk treatments, including response and recovery, and communication plans for the identified and measured AI risks are documented and monitored regularly.
- Manage 4.1: Post-deployment AI system monitoring plans are implemented, including mechanisms for capturing and evaluating input from users and other relevant AI actors, appeal and override, decommissioning, incident response, recovery, and change management.
- Manage 4.2: Measurable activities for continual improvements are integrated into AI system updates and include regular engagement with interested parties, including relevant AI actors.
- Manage 4.3: Incidents and errors are communicated to relevant AI actors including affected communities. Processes for tracking, responding to, and recovering from incidents and errors are followed and documented.